\documentstyle[12pt]{article} 
\input epsf
\begin{document}   
\title{\bf The Moyal Momentum algebra applied to $\theta$-deformed $2d$ conformal models and KdV-hierarchies}
\author{\\A. BOULAHOUAL  and  M. B. SEDRA\footnote{Corresponding author: mysedra@yahoo.fr.}\\
\small{Abdus Salam International Centre For Theoretical Physics ICTP,
Trieste, Italy}\\
\small{and}\\
\small {Laboratoire de Physique Th\'eorique et Appliqu\'ee LPTA,} \\
\small{Universite Ibn Tofail, Facult\'e des Sciences, D\'epartement de Physique,}\\ 
\small{B.P. 133, Kenitra, Morocco}}
\maketitle 
\hoffset=-1cm 
\textwidth=11,5cm                        
\vspace*{1cm} 
\begin{abstract}
The properties of the Das-Popowicz Moyal momentum algebra that we introduce in hep-th/0207242 are reexamined in details and used to discuss some aspects of integrable models and 2d conformal field theories. Among the results presented we setup some useful convention notations which lead to extract some non trivial properties of the Moyal momentum algebra. We use the particular sub-algebra $sl_{n}-{\widehat \Sigma}_{n}^{(0,n)}$ to construct the $sl_2$-Liouville conformal model $\partial{\bar \partial}\phi=\frac{2}{\theta}e^{-\frac{1}{\theta}\phi}$ and its $sl_3$-Toda extension $\partial{\bar \partial}\phi_1 = Ae^{-\frac{1}{2\theta}(\phi_1+\frac{1}{2}\phi_2)}$ and $\partial{\bar \partial}\phi_2 = Be^{-\frac{1}{2\theta}(\phi_1+2\phi_2)}$. We show also that the central charge, a la Feigin-Fuchs, associated to the spin-2 conformal current of the $\theta$-Liouville model is given by $c_{\theta}=(1+24\theta^{2})$. Moreover, the results obtained for the Das-Popowicz Mm algebra are applied to study systematically some properties of the Moyal KdV and Boussinesq hierarchies generalizing some known results. We discuss  also the primarity condition of conformal $w_{\theta}$-currents and interpret this condition as being a dressing gauge symmetry in the Moyal momentum space. Some computations related to the dressing gauge group are explicitly presented. 
\end{abstract}

\newpage
\tableofcontents
\newpage
\section {Introduction}
Two dimensional integrable models  [1] in connection with conformal field theories [2] and their underlying lower $(s\leq 2) $ [3,4,5,6,7] and higher $(s\geq 2)$ [8,9] spin symmetries, occupy, for several years, a central position in various area of research. More particularly, a lot of interest has been paid to $W$-symmetries [5], which are infinite dimensional algebras extending the conformal invariance (Virasoro algebra) by adding to the energy momentum operator $T(z)\equiv W_{2}$, a set of conserved currents $W_{s}(z)$, of conformal spin $s>2$ with some composite operators necessary for the closure of the algebra. \\\\
In the language of $2d$ conformal field theory, the above mentioned currents $W_s$ are taken in general as primary satisfying the OPE [2]
\begin{equation}
T(z)W_{s}(\omega)=\frac{s}{(z-\omega)^{2}}W_{s}(\omega)+\frac{W'_{s}(\omega)}{(z-\omega)},
\end{equation}
or equivalently, 
\begin{equation}
W_s=J^{s}.{\tilde W}_s
\end{equation}
under a general change of coordinate (diffeomorphism) $x  \rightarrow {\tilde x}(x)$ with $J=\frac{\partial \tilde x}{\partial x}$ is the associated Jacobian.\\\\
These $W$-symmetries exhibit among other a non linear structure and are not Lie algebra in the standard way as they incorporate composite fields in their OPE. \\\\
In integrable models these higher spin symmetries appear such that the Virasoro algebra $W_2$ defines the second Hamiltonian structure for the KdV hierarchy [10, 11], $W_3$ for the Boussinesq [12] and $W_{1+\infty}$ for the KP hierarchy [13] and so one. These correspondences are achieved naturally in terms of pseudo-differential Lax operators [14] 
\begin{equation}
{\mathcal L}_{n}=\sum_{j\in Z}u_{n-j}\partial^{j},
\end{equation}
allowing both positive as well as nonlocal powers of the differential $\partial ^{j}$. The fields $u_j$ of arbitrary conformal spin $j$ did not define a primary basis. The construction of primary fields from the $u_j$ one's is originated from the well known covariantization method of Di-Francesco -Itzykson-Zuber (DIZ)[15] showing that the primary $W_j$ fields are given by adequate polynomials of $u_j$ and their k-th derivatives $u_{j}^{(k)}$.	\\\\
More recently there has been a growth in the interest in non-commutative geometry (NCG), which appears in string theory in several ways [16]. Much attention has been paid also to field theories on NC spaces and more specifically Moyal deformed space-time, because of the appearance of such theories as certain limits of string, D-brane and M-theory [17]. Non-commutative field theories emerging from string (membrane) theory  stimulate actually a lot of important questions about the non-commutative integrable systems and how they can be described in terms of star product and Moyal bracket [18-24]. Recall that in the Moyal momentum algebra the ordinary pseudo differential Lax operators eq(3) are naturally replaced by momentum Lax operators 
 \begin{equation}
{\mathcal L}_{n}=\sum_{j\in Z}u_{n-j}\star p^{j},
\end{equation}
satisfying a non-commutative but associative algebra inherited from the star product.\\\\  
The principal focus of this work and of our earlier note [25] is to provide new insights into integrable models and conformal field theories in the non-commutative geometry framework. The key step towards achieving this aim is via the Moyal momentum algebra introduced in [21] and that we call the Das-Popowicz Mm algebra ${\widehat \Sigma}(\theta)$. We will setup the basic lines of this algebra and study its important properties systematically and look for some important implications of this algebra in integrable hierarchies and conformal field theories.\\\\
We present this work as follows: We start in section 2 by a setup of our convention notations with the basic definitions and present is section 3 a systematic study of the Das-Popowicz Moyal momentum algebra and we show among other results that we can use the sub-algebras $sl_n-{\widehat \Sigma}_{n}^{(0,n)}$ to build the $\theta$ extended $sl_2$-Liouville and $sl_3 $-Toda conformal field theories. \\\\
The results obtained for the Das-Popowicz Mm algebra, are applied is section 4 to study some $\theta$ deformed properties of $sl_2$-KdV and $sl_3$-Boussinesq integrable hierarchies. Our contributions to this study consist in extending the results found in [23] by increasing the order of computations a fact which leads us to discover more important properties. As an original result, we build the $\theta$-deformed $sl_3$-Boussinesq hierarchy and derive the associated $\theta$-flows. \\\\
The next step, section 5,  concerns the $\theta$-generalization of some properties, that we presented in an unpublished note [26], based essentially on the idea to use the dressing gauge group of Lorentz scalar momentum operators $K[a]$ to discuss the Moyal DIZ-covariantization of ${sl_n}-{\widehat \Sigma}^{(0,n)}_{n}$ Lax operators and the primarity condition.\\\\
Recall that the standard method for constructing primary fields $w_k$, is based on the well known DIZ covariantization method of Lax differential operators [15]. This method has leads to build in a successful way all the $w_{k}$ conformal primary fields as functions of the old ones $u_k$ by covariantizing the corresponding Lax operators. Actually this method was extended and applied to the Moyal momentum case by the authors of [24] and has leads to build the $\theta$-deformation of the classical $w_k$-algebra. \\\\
Our alternative way consists in searching for the form of the dressing gauge symmetry $\{K[a]\}$ ensuring in one hand the transition 
\begin{equation}
\{u_i\}\stackrel{K[a]}{ \longrightarrow}\{V_i\}
\end{equation}
from different basis of conformal fields belonging to the ring ${\widehat \Sigma}_{n}^{(0,0)}, n\ge 0$ and on the primarity condition on the other hand. In fact, our request of the invariance of $sl_n$ momentum Lax operators under the action of the dressing gauge group makes strong constraints on the gauge parameters $a_i$ as it's explicitly shown in the $sl_3$ and $sl_4$ examples. Once these parameters are well derived, the associated gauge group is then explicitly determined. \\\\
The knowledge of the dressing  gauge symmetry may provides a geometrical interpretation of the primarity condition of conformal fields as being a gauge choice on some orbit that we call the dressing gauge orbit. Elements of this orbit are well defined gauge dressing groups $\{K_{i}[a]\}$ such that each position on this orbit is characterized by a fixed dressing operator $K_{i}$ which fixes in turn the gauge group. \\\\
\section {Basic definitions and convention notations} 
We need first of all to specify the nature of the objects that we will use in this work. The functions often involved in the two dimensional phase-space 
are arbitrary functions which we generally indicate by $f(x,p)$ with coordinates $x$ and $p$.
With respect to this phase space, we have to define the following objects:\\\\
{\bf 1}. The constants $f_0$ defined such that

\begin{equation}
 \partial_x f_0=0= \partial_p {f_0}
\end{equation}\\\\
{\bf 2}. The functions ${u_i}(x,t)$ depending on an infinite set of variables $t_1 =x, t_2,t_3,...,
$
with
\begin{equation}
\partial_{p} u_{i} (x,t)=0 
\end{equation}
The index $i$, stands for the conformal weight of the field $u_i (x,t)$. 
These functions can be considered in the complex language framework as
being the analytic (conformal ) fields of conformal spin $i=1,2,...$. \\\\
{\bf 3}. Others objects that we will use are the ones given by
\begin{equation}
u_i (x,t)\star p^j  
\end {equation}
which are objects of conformal weight (i+j) living on the
non-commutative space parametrized by $\theta$. Through this work, we will use the following 
convention notations $[u_i]= i$, $[\theta] =0$ and  $[p]=[\partial _x]=-[x]=1$ , where the symbol $[\hspace{0,5 cm}]$ stands for the conformal dimension of the used objects.\\\\
{\bf 4}. The star product law defining the multiplication of objects in the non-commutative 
space is shown to
satisfy the following expression 
\begin{equation}
f(x,p)\star g(x,p)= \sum_{s=0}^{\infty} \sum_{i=0}^{s}{\frac{\theta ^s}{s!}} (-)^{i} c _{s}^{i}
(\partial_{x}^{i}\partial_{p}^{s-i}f)(\partial_{x}^{s-i}\partial_{p}^{i}g)
\end{equation}
with $c _{s}^{i}=\frac {s!}{i!(s-i)!} $.\\\\
{\bf 5}. The Moyal bracket is defined as [18]
\begin{equation}
\{f(x,p), g(x,p)\}_ {\theta} =\frac {f \star g - g \star f}{2\theta}
\end{equation}\\\\
{\bf 6}. To distinguish the classical objects from the $\theta$-deformed ones, we consider the following 
convention notations:\\
a)$ \widehat{\Sigma} _{m}^{(r,s)}$:
Denoting the space of of momentum Lax differential operators of conformal spin $m$ and degrees 
$(r,s)$ with $r\leq s$. Typical operators of this space are given by 
\begin{equation}
\sum _{i=r}^{s}u_{m-i}\star p^{i}.
\end{equation}
\\
b)$\widehat{\Sigma} _{m}^{(0,0)}$:
This is the space of the coefficient functions of conformal spin $m$; $ m\in Z$, which may depend on the parameter $\theta$. It coincides 
in the classical limit, $\theta =0$, with the ring of analytic fields involved into the construction of conformal symmetry and $W$-extensions.\\
c)${\widehat\Sigma}_{m}^{(k,k)}$:
Is the space of momentum operators type,
\begin{equation}
u_{m-k}\star p^{k}                                  
\end{equation}\\
d)$\theta$-Residue operation: $\widehat{Res}$
\begin{equation}
\widehat{Res}(f\star p^{-1})=f
\end{equation}
We will show later how this $\theta$-residue operation is related to the classical residue "Res"\\
{\bf 7}. {\bf Some useful formulas}
\begin{equation}
\begin{array}{lcl}
1\star f(x)g(p)&=&f(x)g(p)\\\\
f(x)p^n\star g(x)&=&f(p^{n}\star g)\\\\
p^n\star f(p)g(x)&=&(p^{n}\star g)f\\\\
f(x)\star g(x)h(p)&=&(f\star h)g\\\\
f(x)p\star g(x)p &=& fgp^2+\theta(fg'-f'g)p-(\theta)^{2}f'g\\\\
\end{array}
\end{equation}

\section {The Das-Popowicz Mm algebra} 
We denote this algebra in our convention notation by ${\widehat \Sigma} (\theta)$.
This is the algebra based on arbitrary momentum Lax operators and which
 decomposes as:

\begin{equation}
{\widehat \Sigma} (\theta) = \oplus _{r\leq s} \oplus _{m \in Z}{\widehat \Sigma}_{m}^{(r,s)}
\end{equation}
In order to construct this huge momentum algebra ${\widehat \Sigma}$, we will start first by looking at the important algebraic properties of its sub-algebras.

\subsection{The Ring of analytic functions: $\Sigma ^{(0,0)}$}
At the beginning, we note that its its convenient to  to use the complex notation which consists to define 
the two dimensional Euclidean space parametrized by $z=t+ix$ and ${\bar z}=t-ix$. In this notation the functions $u_{k}(x,t)\equiv u_{k}(z)$. \\\\
It's then convenient to start by introducing the space of these analytic functions of arbitrary conformal spin. This is the space of completely reducible infinite dimensional $so(2)$ Lorentz representation that can be written as 
\begin{equation}
\Sigma ^{(0,0)}=\oplus_{k\in Z}\Sigma ^{(0,0)}_{k}
\end{equation}
where the $\Sigma_{k} ^{(0,0)}$'s are one dimensional $so(2)$ spin $k$ irreducible modules. The upper indices $(0,0)$ carried by the space, $\Sigma ^{(0,0)})$, are special values of general indices $(p,q)$ to be introduced later on.\\\\
The generators of these spaces are given by the spin $k$ analytic fields. They my be viewed as analytic maps $u_{k}$ which associate to each point $z$ on the unit circle, the fields $u_{k}(z)$. For $k\geq 2$, these fields can be thought of as the higher spin currents involved in the construction 
of $w_{\theta}$-algebras. \\\\
As we are interested in the $\theta$-deformation case, we have to add that the spaces $\Sigma ^{(0,0)}_{k}$ are $\theta$-depending and the corresponding $w_{\theta}$-algebra is shown to exhibit new properties related to the $\theta$ parameter and reduces to the standard $w$-algebra once some special limits on the $\theta$ parameters are performed. \\\\
As an example, consider for instance the $w_{\theta}^{3}$-algebra [23] generalizing the Zamolodchikov algebra [8]. The conserved currents of this extended algebra are shown to take the following form
\begin{equation}
\begin{array}{lcl}
w_2&=&u_{2}\\
w_3&=&u_{3}-\theta u_{2}^{'}
\end{array}
\end{equation}
which coincides with the standard case once $\theta =\frac{1}{2}$.\\\\
As in infinite dimensional spaces, elements $\Phi$ of the spin tensor algebra $\Sigma^{(0,0)}(\theta)$ are constructed from the field basis $\{u_k, k\in Z\}$ as follows
\begin{equation}
\Phi = \Sigma_ {k\in}c(k)u_k,
\end{equation}
where only a finite number of the coefficients $c(k)$ is non-vanishing. Next we introduce the following $\theta$-invariant scalar product $\left < , \right >_{\theta}\equiv {\left < , \right >}$ in the tensor algebra $\Sigma ^{(0,0)}$ 
\begin{equation}
\begin{array}{lcl}
<u_k ,u_l >_{\theta}&=&\delta_{k+l,1}\int{dzu_{1-k}\star u_{k}},\\\\
&=&\delta_{k+l,1}\int{dzu_{1-k}.u_{k}}\\\\
&=& <u_k , u_l >
\end{array}
\end{equation}
The $\theta$-invariance of the introduced ``scalar'' product, is due to the fact that $\partial_p u_k =0$. Using the above product, its not difficult to see that the one dimensional subspaces $\Sigma ^{(0,0)}_{k}$ and $\Sigma ^{(0,0)}_{1-k}$ are dual to each other. This leads to splits the tensor algebra $\Sigma ^{(0,0)}$ into two semi in-finite tensor sub-algebras $\Sigma ^{(0,0)}_{+}$ and $\Sigma ^{(0,0)}_{-}$, characterized respectively by positive and negative conformal spin quantum numbers as shown here below
\begin{equation} 
\begin{array}{lcl}
\Sigma ^{(0,0)}_{+}&=&\oplus _{k>0}\Sigma ^{(0,0)}_{k}\\
\Sigma_{-}^{(0,0)}&=&\oplus _{>0}\Sigma ^{(0,0)}_{1-k}\\
\end{array}
\end{equation}
From these equations, we read in particular that $\Sigma ^{(0,0)}_{0}$ is the dual of $\Sigma ^{(0,0)}_{1}$ and if half integers were allowed, the space $\Sigma ^{(0,0)}_{\frac{1}{2}}$ would be self dual with respect to the product (19).\\\\
Another fact is that these product carries a non vanishing conformal spin quantum number since from dimensional arguments , it behaves as a conformal object of weight $\Delta (< , >)=-1$ as we can easily check from equation (19). This property shows that the product $ \left< , \right > $ don't behaves as a true scalar product and we shall then think to introduce the true Lorentz scalar product.\\\ 
Later on, we will show that it is possible to introduce this object that is the combined scalar product $<< , >>$ build out of (19) and a pairing product (see eq.(45)) such that $\Delta (<< , >>)=0$

\subsection{The space ${\widehat \Sigma}_{m}^{(r,s)}$: Basic properties}
To start let's precise that this space contains momentum operators of fixed conformal spin m and degrees (r,s), type
\begin{equation}
{\mathcal {L}}_{m}^{(r,s)}(u)=\sum _{i=r}^{s}u_{m-i}(x)\star p^i,
\end{equation}
These are $\theta$-differentials whose operator character is inherited from the star product law defined as follows:
 
\begin{equation}
f(x,p)\star g(x,p)= \sum_{s=0}^{\infty} \sum_{i=0}^{s}{\frac{\theta ^s}{s!}} (-)^{i} c _{s}^{i}
(\partial_{x}^{i}\partial_{p}^{s-i}f)(\partial_{x}^{s-i}\partial_{p}^{i}g)
\end{equation}
with $c _{s}^{i}=\frac {s!}{i!(s-i)!} $ and $f(x,p)$ are arbitrary functions on the phase space.
Using this relation, it is now important to precise how the momentum operators act on arbitrary functions $f(x,p)$ via the star product.\\

Performing computations based on the relation (22), we find the following $\theta$- Leibnitz rules:
\begin{equation}
p^{n} \star f(x,p) = \sum _{s=0}^{n} \theta ^{s} c_{n}^{s} f^{(s)}(x,p) p^{n-s},   
\end{equation}
and 
\begin{equation}
p^{-n} \star f(x,p) = \sum _{s=0}^{\infty} (-)^{s} \theta ^{s} c_{n+s-1}^{s} f^{(s)}(x,p) p^{-n-s},  
\end{equation}
where $f^{(s)}=\partial_{x}^{s}f$ is the prime derivative. Few examples are given by 
\begin{equation}
\begin{array}{lcl}
1 \star f(x,p)&=&f\\\\
p\star f(x,p)&=&fp+\theta f'\\\\
p^2\star f(x,p)&=&fp^2+2\theta f'p+{\theta}^2 f''\\\\
p^3\star f(x,p)&=&fp^3+3\theta f'p^2+3\theta^2 f''p+{\theta}^3 f'''\\\\                                        \end{array}      
\end{equation}
and
\begin{equation}
\begin{array}{lcl}
p^{-1}\star f(x,p)&=&fp^{-1}-\theta f'p^{-2} +{\theta}^{2} f'' p^{-3}-{\theta}^{3} f''' p^{-4}+...\\\\
p^{-2}\star f(x,p)&=&fp^{-2}-2\theta f'p^{-3} +3{\theta}^{2} f'' p^{-4}-4{\theta}^{3} f''' p^{-5}+...\\\\
p^{-3}\star f(x,p)&=&fp^{-3}-3\theta f'p^{-4} +6{\theta}^{2} f'' p^{-5}-10{\theta}^{3} f''' p^{-6}+...\\\\
\end{array}
\end{equation}
We find also the following expressions for the Moyal bracket:
\begin{equation}
\begin{array}{lcl}
\{p^n, f\}_{\theta} &=&\sum _{s=0}^{n} \theta ^{s-1} c_{n}^{s}\{\frac {1-(-)^s}{2}\} f^{s} p^{n-s},\\\\ 
\{p^{-n}, f\}_{\theta} &=& \sum _{s=0}^{\infty} \theta ^{s-1} c_{s+n-1}^{s}\{\frac {(-)^{s}-1}{2}\} f^{s} p^{-n-s}, 
\end{array}
\end{equation}
These equations don't contribute for even values of $s$ as we can show in the following few examples 
\begin{equation}
\begin{array}{lcl}
\{p, f\}_{\theta} &=&f'\\\\
\{p^2, f\}_{\theta} &=&2f'p\\\\
\{p^3, f\}_{\theta} &=&3f'p^2+{\theta}^2f'''
\end{array}
\end{equation}
and
\begin{equation}
\begin{array}{lcl}
\{p^{-1}, f\}_{\theta} &=&-f'p^{-2}-{\theta}^{2}f'''p^{-4}-...- {\theta}^{2k}f^{(2k+1)}p^{-2k-2}-...\\\\
\{p^{-2}, f\}_{\theta} &=&-2f'p^{-3}-4{\theta}^{2}f'''p^{-5}-...- (2k+2){\theta}^{2k}f^{(2k+1)}p^{-2k-3}-...\\\\
\{p^{-3}, f\}_{\theta} &=&-3f'p^{-4}-10{\theta}^{2}f'''p^{-6}-...- \frac{(2k+3)(2k+2)}{2}{\theta}^{2k}f^{(2k+1)}p^{-2k-4}-...
\end{array}
\end{equation}
We can then simplify eqs.(27) and write
\begin{equation}
\begin{array}{lcl}
\{p^{n}, f\}_{\theta} &=& \sum _{k=0} {\theta}^{2k}f^{(2k+1)}p^{n-2k-1}\\\\
\{p^{-n}, f\}_{\theta} &=&-\sum _{k=0} {\theta}^{2k}c^{2k+1}_{2k+n}f^{(2k+1)}p^{-2k-n-1}\\\\
\end{array}
\end{equation}
with $c^{2k+1}_{2k+n}=\frac{(2k+n)(2k+n-1)...(2k+2)}{(n-1)!}$. Special Moyal brackets are given by
\begin{equation}
\begin{array}{lcl}
\{p, x\}_{\theta} &=&1\\\\
\{p^{-1}, x\}_{\theta} &=&-p^{-2}
\end{array}
\end{equation}
Now, having derived and discussed some important properties of the Leibnitz rules, we can also remark that the momentum operators $p^i$ satisfy the algebra
\begin{equation}
p^n \star p^m =p^{n+m}.
\end{equation}
which ensures the suspected rule
\begin{equation}
\begin{array}{lcl}
p^{n} \star (p^{-n}\star f) &=& f\\\\
(f\star p^{-n} )\star p^{n}&=&f. 
\end{array}
\end{equation}

Note by the way, that one can represent the momentum Lax operators in two different ways in the space ${\widehat \Sigma}_{m}^{(r,s)}$. Besides the one given in (21), one can equivalently write
\begin{equation}
{\tilde{\mathcal  L}}_{m}^{(r,s)}(U)=\sum _{i=r}^{s}p^{i}\star U_{m-i}
\end{equation}
which corresponds to the so called Volterra representation usually used in the derivation framework of the Gelfand-Dickey second Hamiltonian structure. \\

An important algebraic property of the space ${\widehat \Sigma}_{m}^{(r,s)}$ is that it may decomposes into the underlying subspaces as
\begin{equation}
{\widehat \Sigma}_{m}^{(r,s)} =\oplus _{k=r}^{s} {\widehat \Sigma}_{m}^{(k,k)}(\theta)\\
\end{equation}
where ${\widehat \Sigma}_{m}^{(k,k)}$ are unidimensional subspaces containing prototype elements of kind $u_{m-k}\star p^{k}$ or $p^{k}\star U_{m-k}$.
Using the $\theta$-Leibniz rule, we can write, for fixed value of k:
\begin{equation}
{\widehat \Sigma}_{m}^{(k,k)} \equiv \Sigma _{m}^{(k,k)} \oplus \theta
\Sigma _{m}^{(k-1,k-1)} \oplus \theta^{2}{\Sigma}_{m}^{(k-2,k-2)} \oplus...
\end{equation}
where $\Sigma _{m}^{(k,k)}$ is the standard one dimensional sub-space of Laurent series objects $u_{m-k} p^k$ considered also as the $(\theta=0)$-limit of ${\widehat \Sigma}_{m}^{(k,k)}$. \\\\
This property can be summarized as follows
\begin{equation}
\begin{array}{lcl}
{\widehat \Sigma}_{m}^{(r,s)} &=& \oplus _{k=r}^{s} {\widehat \Sigma}_{m}^{(k,k)}(\theta)\\\\
&=&\oplus _{k=r}^{s}\oplus _{l=0}^{k} {\theta}^{l}{\Sigma}_{m}^{(k-l,k-l)}
\end{array}
\end{equation}
Furthermore, the unidimensional subspaces ${\widehat \Sigma}_{m}^{(k,k)}$  can be written formally as 
\begin{equation}
{\widehat \Sigma}_{m}^{(k,k)} \equiv p^{k} \star \Sigma _{m-k}^{(0,0)}.
\end{equation} 
where ${\widehat \Sigma}_{m}^{(0,0)} \equiv \Sigma _{m}^{(0,0)}$ is nothing but the ring of analytic fields $u_{m}$ of conformal spin ${m \in Z}$ satisfying
\begin{equation}
u_i\star u_j=u_{i}. u_{j}
\end{equation} 
Another property concerning the space ${\widehat \Sigma}_{m}^{(r,s)}$ is its non closure under the action of the Moyal bracket since we have;
\begin{equation}
\{.,.\}_\theta : {\widehat \Sigma}_{m}^{(r,s)} \star {\widehat \Sigma}_{m}^{(r,s)} \rightarrow{\widehat \Sigma}_{2m}^{(r,2s-1)}
\end{equation} 
Imposing the closure, one gets strong constraints on the integers  $m$, $r$ and $s$ namely 
\begin{equation}
\begin{array}{lcl}
m&=&0\\  
r \leq& s& \leq 1
\end{array}  
\end{equation}
With these constraint equations, the sub-spaces ${\widehat \Sigma}_{m}^{(r,s)}$ exhibit then a Lie algebra structure since the $\star$-product is associative. \\\\
The sub-space ${\widehat \Sigma}_{m}^{(r,s)}$ is characterized by the existence of a residue operation that we denote as  ${\widehat Res}$ and which acts as follows
\begin{equation}
\begin{array}{lcl}
\widehat{Res}(u_{k}\star p^{-k}) &=& (u_{k} \star p^{-k}) \delta _{k-1,0}\\\\
&=& u_{1}\delta _{k-1,0}
\end{array}
\end{equation}
This result coincides with the standard residue operation: $Res$, acting on the sub-space   ${\Sigma}_{m}^{(r,s)}$:
\begin{equation}
Res (u_1 . p^{-1})=u_1
\end{equation}
We thus have two type of residues $\widehat Res$ and $Res$ acting on two different spaces ${\widehat \Sigma}_{m}^{(r,s)}$ and  ${\Sigma}_{m}^{(r,s)}$ but with value on the same ring ${\Sigma}_{m+1}^{(0,0)}$. This Property is summarized as follows:
\begin{equation}
\begin{array}{lcl}

{\widehat \Sigma}_{m}^{(r,s)}  \hspace{0,5cm}\stackrel {\small\theta =0} {\longrightarrow} \hspace{0,5cm}  {\Sigma _{m}^{(r,s)}} \\

\hspace{0,2cm}{_{\small \widehat Res}} {\large\searrow}   \hspace{1cm}  {\large\swarrow}_{\small {Res}}   \\  
\hspace{1,15cm} \Sigma _{m+1}^{(0,0)} 
\end{array}
\end{equation}

We learn from this diagram that the residue operation exhibits a conformal spin quantum number equal to 1.\\
With respect to the previous residue operation, we define on ${\widehat \Sigma}$ the following degrees pairing product 
\begin{equation}
\left(.,.\right): {\widehat \Sigma}_{m}^{(r,s)} \star {\widehat \Sigma}_{n}^{(-s-1,-r-1)} \rightarrow  {\Sigma}_{m+n+1}^{(0,0)}          
\end{equation}
such that
\begin{equation}
\left({\mathcal L }_{m}^{(r,s)}(u),\tilde {\mathcal L }_{n}^{(\alpha ,\beta)}(v)\right)=\delta_{\alpha +s+1,0}\delta_{\beta+r+1,0}{\widehat Res}\left[{\mathcal L }_{m}^{(r,s)}(u)\star \tilde {\mathcal L }_{n}^{(\alpha ,\beta)}(v)\right],
\end{equation}
showing that the spaces ${\widehat \Sigma}_{m}^{(r,s)}$ and  ${\widehat \Sigma }_{n}^{(-s-1 ,-r-1)}$ are $\widehat Res$-dual as ${\Sigma}_{m}^{(r,s)}$ and  ${\Sigma }_{n}^{(-s-1 ,-r-1)}$ are dual with respect to the $Res$-operation.\\ 

\subsection{The huge Lie algebra ${\widehat \Sigma}_{0}^{(-\infty,1)}$}
As discussed previously, the subspaces ${\widehat \Sigma}_{m}^{(r,s)}$  exhibit a Lie algebra structure with respect to the Moyal bracket once the spin-degrees constraints eq.(41) are considered. With these conditions one should note that the huge Lie algebra that we can extract from the space ${\widehat \Sigma}_{m}^{(r,s)}$ consists on the space ${\widehat \Sigma}_{0}^{(-\infty,1)}$ having the remarkable space decomposition
\begin{equation}
{\widehat \Sigma}_{0}^{(-\infty,1)}={\widehat \Sigma}_{0}^{(-\infty,-1)}\oplus {\widehat \Sigma}_{0}^{(0,1)},
\end{equation}
where ${\widehat \Sigma}_{0}^{(-\infty,-1)}$ describes the Lie algebra of pure non local momentum operators and ${\widehat \Sigma}_{0}^{(0,1)}$ is the Lie algebra of local Lorentz scalar momentum operators ${\mathcal L}_{0}(u)= u_{-1}\star p+u_0$. The latter can splits as follows
\begin{equation}
{\widehat \Sigma}_{0}^{(0,1)}={\widehat \Sigma}_{0}^{(0,0)}\oplus {\widehat \Sigma}_{0}^{(1,1)},
\end{equation}
where ${\widehat \Sigma}_{0}^{(1,1)}$ is the Lie algebra of vector momentum fields $J_{0}(u)=u_{-1}\star p$ which are also  elements of ${\Sigma}_{0}^{(0,1)}$.\\\\
As a prototype example, consider
\begin{equation}
\begin{array}{lcl}
{\mathcal L}_{u}&=& u_{-1}\star p+u_0  \\\\
{\mathcal L}_{v}&=& v_{-1}\star p+v_0
\end{array}
\end{equation}
be two elements of  ${\widehat \Sigma}_{0}^{(0,1)}$. Straightforward computations lead to the following Moyal bracket algebra;
\begin{equation}
\{{\mathcal L}_{u},{\mathcal L}_{v} \}_{\theta}={\mathcal L}_{w}
\end{equation}
with 
\begin{equation}
\begin{array}{lcl}
{\mathcal L}_{w}&=& w_{-1}\star p+w_0\\\\
&=& \{u_{-1}v'_{-1}-u'_{-1}v_{-1}\}\star p +\{u_{-1}v'_{0}-u'_{0}v_{-1}\}.
\end{array}
\end{equation}
Forgetting about the fields ({\it of vanishing conformal spin}) belonging to ${\Sigma}_{0}^{(0,0)}$ is equivalent to consider the coset space
\begin{equation}
{\widehat \Sigma}_{0}^{(1,1)}\equiv  {{\widehat \Sigma}_{0}^{(0,1)}}/ {{\Sigma} _{0}^{(0,0)} }
\end{equation}
one obtain the $Diff(S^1)$ momentum algebra of vector fields $J_{0}(u)=u_{-1}\star p$ namely
\begin{equation}
\{J_{0}(u), J_{0}(v)\}_{\theta}=J_{0}(w)
\end{equation}
with $w_{-1}=u_{-1}v'_{-1}-u'_{-1}v_{-1}$.\\

The extension of these results to non local momentum operators is natural. In fact, one easily show that the previous Lie algebras are simply sub-algebras of the huge momentum space ${\widehat \Sigma}_{0}^{(-\infty,1)}$. For a given $0\leq k\leq 1$, we have
\begin{equation}
{\widehat \Sigma}_{0}^{(0, 1)}\subset {\widehat \Sigma}_{0}^{(-\infty, k)}\subset {\widehat \Sigma}_{0}^{(-\infty, 1)}
\end{equation}
and by virtue of (47) 
\begin{equation}
\{{\widehat \Sigma}_{0}^{(-\infty,k)},  {\widehat \Sigma}_{0}^{(0, 1)}\}_{\theta}\subset {\widehat \Sigma}_{0}^{(-\infty, k)}\subset {\widehat \Sigma}_{0}^{(-\infty,1)}
\end{equation}
and for $-\infty < p \leq q\leq 1$
\begin{equation}
\{{\widehat \Sigma}_{0}^{(-\infty, p)},  {\widehat \Sigma}_{0}^{(-\infty, q)}\}_{\theta}\subset {\widehat \Sigma}_{0}^{(-\infty, p+q-1)}
\end{equation}
These Moyal bracket expressions show in turn that all the subspaces ${\widehat \Sigma}_{0}^{(p,q)}$ with  $-\infty < p \leq q\leq 1$ are ideals of of ${\widehat \Sigma}_{0}^{(-\infty,1)}$.

\subsection{The huge space ${\widehat \Sigma}$}
This is the algebra of momentum operators of arbitrary conformal spin and arbitrary degrees. It is obtained by summing over all allowed values of spin and degrees in the following way 
\begin{equation}
\begin{array}{lcl}
{\widehat \Sigma}&=&\oplus_{p\leq q}{\widehat \Sigma}^{(p,q)} \\\\
&=& \oplus_{p\leq q}\oplus_{m\in Z}{\widehat \Sigma}_{m}^{(p,q)}.
\end{array}
\end{equation}
This infinite dimensional momentum algebra is closed under the Moyal bracket without any condition. A remarkable property of this space is the possibility to introduce six infinite dimensional classes of momentum sub-algebras related to each others by special duality relations that we will precise. These classes of algebras are given by ${\widehat \Sigma}_{s}^{\pm},$ with $s=0,+,-$ describing respectively the different values of the conformal spin which can be zero, positive or negative. The $\pm$ upper indices stand for the values of the degrees quantum numbers. \\\\
As we explicitly show in sec.(3.2) the degrees pairing that we introduce eqs.(45-46) is very useful as it leads to consider the subspaces ${\widehat \Sigma}_{m}^{(r,s)}$ and  ${\widehat \Sigma }_{n}^{(-s-1 ,-r-1)}$ as $\widehat Res$-duals to each others.
Now with this degrees-pairing and the ``scalar' product eq.(19) carrying a conformal dimension equal to $(-1)$, one can introduce a combined scalar product $<< , >> $ carrying the true dimension namely $[<< , >>]=0$, such that for given ${\widehat Res}$-dual momentum operators ${\mathcal L }_{m}^{(r,s)}(u)$ and $\tilde {\mathcal L }_{n}^{(\alpha ,\beta)}(v)$ one can consider the following definition,\\
\begin{equation}
\left<\left<{\mathcal L }_{m}^{(r,s)}(u),\tilde {\mathcal L }_{n}^{(\alpha ,\beta)}(v) \right>\right> =\delta_{m+n,0} \delta_{\alpha +s+1,0}\delta_{\beta+r+1,0}\int{dx {\widehat Res}\left[{\mathcal L }_{m}^{(r,s)}(u)\star \tilde {\mathcal L }_{n}^{(\alpha ,\beta)}(v)\right] }
\end{equation}\\
showing that the spaces ${\widehat \Sigma}_{m}^{(r,s)}$ and  ${\widehat \Sigma }_{-m}^{(-s-1 ,-r-1)}$ are $\widehat Res$-dual as do the spaces  ${\Sigma }_{m}^{(r ,s)}$ and  ${\Sigma }_{-m}^{(-s-1 ,-r-1)}$ in the classical case [7]. \\\\
To illustrate this $\widehat Res$-duality, let's consider the following example 
\begin{equation}
\left(.,.\right): {\widehat \Sigma}_{2}^{(0,2)} \star {\widehat \Sigma}_{-2}^{(-3,-1)} \rightarrow  {\Sigma}_{1}^{(0,0)}\end{equation}
with 
\begin{equation}
\begin{array}{lcl}
{\mathcal L }_{2}^{(0,2)}&=& {u_{0}\star p}^{2} +u_{1}\star p +u_2 \\\\
&=&u_{0}{p}^{2} +(u_{1}-2\theta u'_{0})p +(u_{2}-\theta u'_{1}+\theta^{2} u''_{0})
\end{array}
\end{equation}
and 
\begin{equation}
\begin{array}{lcl}
\tilde {\mathcal L }_{-2}^{(-3,-1)}&=& {v_{1}\star p}^{-3} +v_{0}\star p^{-2} +v_{-1}\star p^{-1}\\\\
&=& v_{-1}{p}^{-1} +(v_{0}+\theta v'_{-1})p^{-2} +(v_{1}+2\theta v'_{0}+\theta^{2} u''_{-1})p^{-3}
\end{array}
\end{equation}
One can easily check that these momentum operators are $\widehat Res$-duals to each others. In fact, computing ${\widehat Res}\{{\mathcal L}\star \tilde {\mathcal L}\}$ one find the following result
\begin{equation}
\begin{array}{lcl}
{\widehat Res}\{ {\mathcal L}_2 \star \tilde{\mathcal L}_{-2} \}&=&(u_{0}v_{1}+ u_{1}v_{0}+u_{2}v_{-1})\\\\
&+& 2\theta (u_{1}v'_{-1}+ 2u_{0}v'_{0})\\\\
&+& 4u_{0}v''_{-1}{\theta}^{2}, 
\end{array}
\end{equation}
showing that ${\mathcal L}(u_0, u_1, u_2)$ is the ${\widehat Res}$-dual of $\tilde {\mathcal L}(v_{1}, v_{0}, v_{-1})$. We learn also that ${\widehat Res}\{{\mathcal L} \star \tilde {\mathcal L}\}$ is an object of conformal spin 1. \\\\
Since we usually require for ${\mathcal L}(u_0, u_1, u_2)$ to be an $sl_{2}$-Lax momentum operator, we may set 
\begin{equation}
\begin{array}{lcl}
u_{0}&=&1\\
u_{1}&=&0
\end{array}
\end{equation}
These conditions on the space ${\widehat \Sigma}^{(0,2)}_{2}$ will respectively induce two dual constraints on the space ${\widehat \Sigma}^{(-3,-1)}_{-2}$ namely 
\begin{equation}
\begin{array}{lcl}
v_{1}&=&0\\\\
{\widehat Res}\{ {\mathcal L }_{2}, \tilde {\mathcal L}_{-2}\}_{\theta}&=&0
\end{array}
\end{equation}
The latter equation is nothing but the constraint which leads to fix the value of the field $v_0$, the $\widehat Res$-dual of of $u_1=0$. Computing this Moyal bracket for this simple example we find 
\begin{equation}
v_0=-\theta v'_{-1}
\end{equation}
The $sl_2$ constraint equations mean also that the real $\widehat Res$-dual of ${\widehat \Sigma}^{(0,2)}_{2}$ is ${\widehat \Sigma}^{(-2,-1)}_{-2}$ as the lower term in $\tilde {\mathcal L}_{-2}$ namely $v_{1}\star p^{-3}$ does not contribute.

For the second example namely: $sl_3$,  we have 
\begin{equation}
\begin{array}{lcl}
{\mathcal L }_{3}^{(0,3)}&=& p^{3} +u_{2}p+(u_3 -\theta u'_{2})\\\\
{\tilde {\mathcal L }}_{-3}^{(-3,-1)}&=& {v_{0}\star p}^{-3} +v_{-1}\star p^{-2} +v_{-2}\star p^{-1}\\\\
\end{array}
\end{equation}
These momentum operators are $\widehat Res$-duals to each others. In fact, once again, explicit computations give
\begin{equation}
{\widehat Res} ({\mathcal L}_3 \star {\tilde {\mathcal L}_{-3}} )=3v'_{0}+u'_{-2}v_{-2}+ u_{-2}v'_{-2}+4\theta^{2}v'''_{-2}+6\theta v''_{-1}
\end{equation}
The field $v_0$ is subject to the following constraint on the Moyal bracket
\begin{equation}
{\widehat Res}\{ {\mathcal L }_{3}, \tilde {\mathcal L}_{-3}\}_{\theta}=0
\end{equation}
which gives 
\begin{equation}
v_{{0}} =-\frac{1}{3} (u_{2}v_{-2}+6\theta v'_{-1}+4{\theta}^{2} v''_{-2})
\end{equation}    
Note finally that we can generalize this analysis to build the $\widehat Res$-dual for Lax momentum operators of higher degrees. Requiring the $sl_n$ symmetry, one show the following result
\begin{equation}
{\widehat Res}({\widehat \Sigma}^{(0,n)}_{n})\equiv {\widehat \Sigma}^{(-n,-1)}_{-n}
\end{equation}
and the $\widehat Res$-dual of the field $u_1=0$ namely $v_0$ is obtained once we solve the general constraint equation
\begin{equation}
{\widehat Res}\{ {\mathcal L }_{n}, \tilde {\mathcal L}_{-n}\}_{\theta}=0
\end{equation}
\subsection{The space $sl_n-\widehat \Sigma_{n}^{(0,n)}(\theta)$ and $2d$ conformal field theory}
This is simply the coset space ${\widehat \Sigma}_{n}^{(0,n)}/{\widehat \Sigma}_{n}^{(1,1)}$ of $sl_n$-Lax operators given by
\begin{equation}
{{\mathcal L}_{n}}(u)=p^n + \sum _{i=0}^{n-2}u_{n-i}\star p^i
\end{equation}
where we have set $u_0 = 1 $ and $u_1 = 0$. This is a natural generalization of the well known $sl_{2}$-momentum Lax operator 
\begin{equation}
{\mathcal L }_{2}= {p}^{2} +u_2
\end{equation}
associated to the $\theta$-KdV integrable hierarchy that we will discuss later.\\\\ 
$sl_n$-momentum Lax operators play a central role in the study of integrable models and more particularly in deriving higher conformal spin algebras ($w_{\theta}$-algebras) from the $\theta$-extended Gelfand-Dickey second Hamiltonian structure [23]. Since they are also important in recovering $2d$ conformal field theories via the Miura transformation, we guess that its possible to extend this property, in a natural way, to the non-commutative case and consider the $\theta$-deformed analogue of the well known $2d$ conformal models namely: the $sl_{2}$-Liouville field theory and its $sl_n$-Toda extensions and also the Wess-Zumino- Novikov-Witten conformal model.\\\\
As an example consider the  $\theta$-KdV momentum Lax operator that we can write as 
\begin{equation}
\begin{array}{lcl}
{\mathcal L }_{2}&=& {p}^{2} +u_2\\
&=&(p+{\phi}')\star(p-{\phi}')
\end{array}
\end{equation}
where $\phi$ is a Lorentz scalar field. As a result we have 
\begin{equation}
u_2= -{\phi '}^{2}-2{\theta}{\phi}''
\end{equation}
which is nothing but the $\theta$ analogue of classical stress energy momentum tensor of $2d$ conformal Liouville field theory. Using $2d$ complex coordinates language, we can write
\begin{equation}
T_{\theta}(z)\equiv u_2(z)= -2\theta{\partial^{2}}\phi -(\partial \phi)^2
\end{equation}
with $\partial \equiv \partial_{z}\equiv \frac{\partial}{\partial z} $. The conservation for this conformal current namely $\bar{\partial} T(z)=0$, leads to write the following $\theta$-Liouville equation of motion
\begin{equation}
\partial{\bar \partial}\phi=\frac{2}{\theta}e^{-\frac{1}{\theta}\phi}
\end{equation}
associated to the two dimensional $\theta$-Liouville action
\begin{equation}
S=\int {d^{2}z\left(\frac{1}{2}{\partial \phi}\star{\bar \partial \phi}+e^{-\frac{1}{\theta}\phi} \right)}
\end{equation}
with ${\partial \phi}\star{\bar \partial \phi}={\partial \phi}{\bar \partial \phi}$.\\\\
Note by the way that  we may interpret the inverse $\frac{1}{\theta}$ of the non-commutative $\theta$-parameter as being the analogue of the Cartan matrix of $sl_2$ because in the classical limit this Cartan matrix is known to be $(a_{ij})=2$\\
Another important point is that we know from the standard $2d$ CFT [2] that the object $T_{zz}$ satisfying 
\begin{equation}
T_{zz}= -\frac{1}{4}(\partial \phi)^2+i\alpha_{0}{\partial^{2}}\phi 
\end{equation}
is nothing but the Feigin-Fuchs representation of the conserved current generating the conformal invariance of a quantum conformal model with the central charge $c=(1-24{\alpha_0}^2)$.\\\\
Using this standard result, we can conclude that the $\theta$-conformal current that we derive in eq(76) (with the rescaling $\phi\equiv \frac{1}{2}\phi$) is associated to the $\theta$-Liouville model having the central charge  
\begin{equation}
c_{\theta}=(1+24{\theta}^2)
\end{equation}
where the non-commutative parameter is shown to coincide with $\alpha_0$ as follows $\theta=-i\alpha_0$.\\\\
The analysis that we use to derive the $\theta$-Liouville equation and its central charge a la Feigin-Fuchs, can be generalized to higher conformal spin Toda field theories associated to $sl_n$ symmetry with $(n-1)-$conserved currents $T(z), w_3, w_4,...w_n$. \\\\
The first non trivial $\theta$-deformed Toda field theory is the one associated to the $sl_3$ momentum Lax operator
\begin{equation}
\begin{array}{lcl}
{\mathcal L}_{3}=p^3+u_2p+w_3
\end{array}
\end{equation}
with $w_3=u_3-\theta u'_2$. It reads in the Miura transformation as
\begin{equation}
{\mathcal L }_{3}=(p+{\partial \phi_1})\star(p+{\partial \phi_2})\star(p-{\partial \phi_1}-{\partial \phi_2})
\end{equation}
where $\phi_k$ are Lorentz scalar fields s.t. $[\phi_k]=0$. Identifying the two expressions of ${\mathcal L}_{3}$, we obtain\\
\begin{equation}
\begin{array}{lcl}
u_2&=& -2\theta({\partial}^{2}\phi_{2}+2{\partial}^{2}{ \phi}_{1})-
({\partial \phi_2}^{2}+{\partial \phi_1}^{2}+{\partial \phi_1}{\partial \phi_2})\\\\
w_3&=&-2{\theta}^{2}\partial^{3} \phi_{2}-\theta({\partial^{2} \phi_{1}}{\partial \phi_{2}}+
3{\partial \phi_{1}}{\partial^{2} \phi_{2}}+2{\partial\phi_{2}}{\partial^{2} \phi_{2}})-
{\partial\phi_{1}}^{2}{\partial \phi_{2}}-{\partial\phi_{1}}{\partial \phi_{2}}^2.
\end{array}
\end{equation}\\
Now its important to look for the associated $sl_3$-Toda equations of motion. 
This is a set of two independent equations having the following form in the (commutative) standard case.\\
\begin{equation}
a\partial{\bar \partial}\phi_i=be^{\sum_{j=1}^{2}K_{ij}\phi_j}
\end{equation}\\
where $K_{ij}$ is the Cartan matrix of $sl_3$ of rank 2 and where $a,b$ are arbitrary constants that can be choice in a convenient way.\\
Performing the same techniques used in deriving the $\theta$-Liouville equation of motion eq.(77), we find the following $\theta$-deformed $sl_3$-Toda equations 
\begin{equation}
\begin{array}{lcl}
\partial{\bar \partial}\phi_1&=&Ae^{-\frac{1}{2\theta}(\phi_1+\frac{1}{2}\phi_2)}\\\\
\partial{\bar \partial}\phi_2&=&Be^{-\frac{1}{2\theta}(\phi_1+2\phi_2)}
\end{array}
\end{equation}
These equations are compatible with the conservation of the conformal spin two current eq(83).\\\\
Once again one can see the Cartan matrix of $sl_3$ as being a $\theta^{-1}$ dependent object. In fact, from the previous established Toda equations of motion, the $\theta$-deformed Cartan matrix of $sl_3$ $K^{\theta}_{ij}\equiv (K^{\theta}_{11}, K^{\theta}_{12}, K^{\theta}_{21}, K^{\theta}_{22})$  is shown to take the following form
\begin{equation}
K^{\theta}_{ij}\equiv -\frac{1}{2\theta} \left(\matrix{1&\frac{1}{2} \cr 1 &2\cr}\right)
\end{equation}
Finally we note that all the properties discussed above may be generalized to the $sl_n$ case. This is an explicit proof of the importance of the algebraic structure inherited from the Das-Popowicz Moyal momentum algebra. Actually we showed how these algebra may leads to extend, in a successful, way all the important properties of $2d$ CFT theories. We will present in the next section some other applications of the momentum algebra in $\theta$-integrable KdV hierarchies

\section {$sl_n$-Moyal KdV-hierarchy}
The aim of this section is to present some results related to the $\theta$-KdV hierarchy. Using our convention notations [25] and the analysis that we developed previously, we will perform hard algebraic computations and derive the $\theta$-KdV hierarchy. The obtained results are shown to be compatible with those presented in [23].\\\\
These computations are very hard and difficult to realize in the general case.  We will simplify this study by limiting our computations to the first orders of the hierarchy namely the $sl_2$-KdV and $sl_3$-Boussinesq $\theta$-integrable hierarchies.\\\\ 
Our contribution to this study consists in extending the results found in [23] by increasing the order of computations a fact which leads us to discover more important properties as we will explicitly show. As an original result, we will build the $\theta$-deformed $sl_3$-Boussinesq hierarchy and derive the associated $\theta$-flows. Some other important results are also presented.  
\subsection{$sl_2$-KdV hierarchy}
Let's consider the $sl_2$-momentum Lax operator 
\begin{equation}
{\mathcal L}_2= p^2 + u_2
\end{equation}
whose 2th root is given by  
\begin{equation}
\begin{array}{lcl}
{\mathcal L}^{\frac{1}{2}}&=& \Sigma_{i=-1} b_{i+1} \star p^{-i}\\\\
&=&{\Sigma}_{i=-1} a_{i+1}p^{-i} 
\end{array}
\end{equation}
This 2th root of ${\mathcal L}_2$ is an object of conformal spin $[{\mathcal L}^{\frac{1}{2}}]=1$ that plays a central role in the derivation of the $\theta$-Lax evolutions equations. In the spirit to contribute much more to this $sl_2$-KdV hierarchy, it was important for us to recover the already established results [23]. \\\\
Performing lengthy but straightforward calculations we compute the coefficients $b_{i+1}$ of ${\mathcal L}^{\frac{1}{2}}$ up to $i=7$ given by \footnote{The authors of [23] present explicit computations of the coefficients $a_{i+1}$ and omit the $b_{i+1}$ ones. Here, we give explicit computation of both of them}:   
\begin{equation}
\begin{array}{lcl}
b_0 &=& 1 \\\\
b_1 &=& 0 \\\\
b_2 &=&\frac{1}{2} u \\\\
b_3&=&-\frac{1}{2} \theta u^{'} \\\\
b_4&=&-\frac{1}{8}u^{2}+\frac{1}{2}\theta^{2} u^{''} \\\\
b_5&=&-\frac{1}{2}\theta^3 u^{'''}+\frac{3}{4}\theta u u^{'}\\\\ 
b_6&=&\frac{1}{16}u^{3}-\frac{7}{4}\theta^{2}u u^{''}-\frac {11}{8}{\theta}^{2} (u^{'})^{2}+\frac{1}{2}{\theta}^{4} u^{''''}\\\\
b_7&=&-\frac{15}{16}\theta u^{2}u'+{\theta}^{3}(\frac{15}{2}u''u'+\frac{15}{4}uu'''')-\frac{1}{2}{\theta}^{5}u^{(5)}\\\\
b_8&=&-\frac{5}{128} u^{4}+{\theta}^{2}(\frac {55}{16}u'' u^{2} +\frac {85}{16}u {u'}^2)-{\theta}^{4}(\frac {31}{4}u u'''' +\frac {91}{8}u''^{2}+\frac {37}{2}u' u''')+\frac {1}{2}{\theta}^{6} u^{(6)}
\end{array}
\end{equation}
and 
\begin{equation}
\begin{array}{lcl}
b_{9}&=&\frac{35}{32}\theta u^{3}u'-\frac{175}{4}{\theta}^{3}\left( u u'u''+\frac{1}{4}u'^3+\frac{1}{4}u^{2}u''' \right)+\frac{7}{4}{\theta}^{5}\left(9u u^{(5)}+25u^{(4)}u'+35u'''u''\right)\\\\
& &- \frac{1}{2}\theta^{7}u^{(7)}\\\\
b_{10}&=&{\frac {7}{256}}u^{5}-\frac {35}{32} {\theta}^{2}\left(\frac{23}{2}u^{2}u^{'2}+5u^{3}u''\right)+\frac {7}{4} {\theta}^{4}\left(\frac{73}{4}u^{2}u^{(4)}+\frac{227}{4}uu^{''2}
+\frac{337}{4}u''u^{'2}+89uu'u'''\right)\\\\
& &-\frac {3}{4} {\theta}^{6}\left(\frac{631}{3}u^{''}u^{(4)}+233u{u'''}^{2}+135u'u^{(5)}\right)+\frac{1}{2}\theta^{8}u^{(8)}\\\\
\end{array}
\end{equation}
These results are obtained by using the identification $ {\mathcal L}_{2}={\mathcal L}^{\frac{1}{2}}*{\mathcal L}^{\frac{1}{2}}$. Note that by virtue of eq.(88), the coefficients $a_{i+1}$ are shown to be functions of $b_{i+1}$ and their derivatives in the following way
\begin{equation}
a_{i+1}=\sum_{s=0}^{i-1}{\theta^{s}} c^{s}_{i-1} b^{(s)}_{i+1-s},
\end{equation}
Substituting the derived expressions of $b_{i+1}$ eqs.(89-90) into eq.(91), we obtain the results presented in [23] namely:\footnote{In [23], important explicit computations of the parameters $a_{i+1}$ are presented up to $a_{10}$. Our calculus eq(92-93), performed up to $a_{12}$ show some missing terms in the computations of [23] relative to $a_{10}$}
\begin{equation} 
\begin{array}{lcl}
a_0 &=& 1 \\\\
a_2 &=&\frac{1}{2} u \\\\
a_4&=&-\frac{1}{8}u^{2} \\\\
a_6&=&\frac{1}{16}u^{3}+\frac{1}{8}{\theta }^{2}(u'^{2}-2u u^{''})\\\\
a_{8}&=&-{\frac {5}{128}}u^{4}+{\frac {5}{8}}{\theta}^{2} \left( u ^{2}u''- \frac{1}{2} u ^{'2}u \right)+{\frac {1}{4}}{\theta}^{4} \left( u''' u'- uu^{(4)}-\frac{1}{2} u ^{''2}\right)\\\\
a_{10}&=&{\frac {7}{256}}u^{5}+{\frac {35}{64}}{\theta}^{2} \left( \frac{1}{2} u ^{2}u^{'2}- u^{3}u''\right)+{\frac {7}{4}}{\theta}^{4} \left( \frac{3}{4} u^{(4)}u^2+\frac{7}{4} u^{''2}u-\frac{3}{4}u'^{2}u''-uu' u'''\right)\\\\
&+&{\frac {1}{4}}{\theta}^{6} \left(u'u^{(5)}+\frac{1}{2} u'''^{2}-uu^6 \right)\\\\
a_{12}&=&-{\frac {21}{1024}}u^{6}+{\frac {105}{64}}{\theta}^{2} \left(u^{4}u''-\frac{1}{2} u^{3}u'^{2}\right)\\\\
&+&{\frac {1}{16}}{\theta}^{4} \left(147uu''u'^{2}+ \frac{189}{2} u^{2}u'u'''-\frac{1029}{4} u^{2}u^{''2}-63u^{3}u^{(4)}-\frac{105}{8} u'^{4} \right)\\\\
&+&{\frac {1}{4}}{\theta}^{6} \left(16u^{''3}+9u^{2}u^{(6)}- 27u'u''u'''-\frac{45}{2} u'^{2}u^{(4)}-\frac{69}{4} u^{'''2}u+\frac{153}{2} uu''u^{(4)}-\frac{27}{2} uu'u^{(5)}\right)\\\\
&+&{\frac {1}{4}}{\theta}^{8} \left(u'u^{(7)}+u'''u^{(5)}-u''u^{(6)}-uu^{(8)}-\frac{1}{2} {u^{(4)}}^2\right)\\& &\vdots
\end{array}
\end{equation}
with 
\begin{equation}
a_{2k+1}=\sum_{s=0}^{2k-1}{\theta^{s}} c^{s}_{2k-1} b^{(s)}_{2k+1-s}=0,\hspace{1cm} k=0,1,2,3,...
\end{equation}
Now having derived the explicit expression of ${\mathcal L}^{\frac{1}{2}}$, we are now in position to write the explicit forms of the set of $sl_n$-Moyal KdV hierarchy. These equations defined as
\begin{equation}
\frac{\partial {\mathcal L}}{\partial t_{k}}=\{({\mathcal L}^{\frac{k}{2}})_{+} ,{\mathcal L}\}_{\theta},
\end{equation}\\
are computed in [23] up to the first three flows $t_{1}, t_{3}, t_{5}$. We work out these equations by adding other flows namely $t_{7}$ and $t_{9}$. We find
\begin{equation}
\begin{array}{lcl}
{u}_{t_{1}}&=&{u'}\\\\
{u}_{t_{3}}&=&\frac{3}{2}uu'+{\theta^2}u'''\\\\
{u}_{t_{5}}&=&\frac{15}{8}u^{2}u'+5{\theta^2}(u'u''+\frac{1}{2}uu''')+{\theta}^{4}u^{(5)}\\\\
{u}_{t_{7}}&=&{\frac{35}{16}} u^{3}u^{\prime }+\frac{35}{8}{\theta }%
^{2} ( {4} uu^{\prime }u^{\prime \prime }+ u^{\prime
3}+ u^{2}u^{\prime \prime \prime } ) +\frac{7}{2}(
uu^{(5)}+3 u^{\prime }u^{(4)}+{5} u^{\prime \prime }u^{\prime \prime
\prime } ) {\theta }^{4}+{\theta }^{6}u^{(7)}\\\\
{u}_{t_{9}}&=& 18{\theta}^{6}u'u^{(6)} +{\frac {651}{8}}{\theta}
^{4} 
u'(u'')^{2}+{\frac {315}{128}}u^{4}u' +{\frac {483}{8}}{\theta}^{4} 
u'^{2}u'''+{\frac {315}{16}}{\theta}^{2}uu'^{3}+{\frac {189}{4}}{
\theta}^{4}uu^{(4)}u' \\\\ 
&+&{\frac {
315}{8}}{\theta}^{2}u^{2}u'u''+{\frac {315}{4}}{\theta}^{4}u  
u'u''' +63{\theta}^{6}u'''u^{(4)} +\frac {105}{16}\theta^{2}
u^{3}u''' +42{\theta}^{6}u^{(5)}u''\\\\ 
&+&{\frac {63}{8}}{\theta}^{4}u^{2}u^{(5)}+{\theta}^{8}
u^{(9)}+\frac{9}{2}{\theta}^{6}uu^{(7)}
\end{array}
\end{equation}
\\\\Ssome important remarks are in order:\\\\
{\bf 1}. The flow parameters $t_{2k+1}$ has the following conformal dimension $[\partial_{t _{2k+1}}]=-
[t_{2k+1}]=2k+1$ for $k=0,1,2,...,$.\\\\
{\bf 2}. A remarkable property of the $sl_2$-Moyal KdV hierarchy is about the degree of non linearity of the $\theta$-evolution equations (95). We present in the following table the behavior of the higher non-linear terms with respect to the first leading flows $t_{1},...,t_{9}$ and give the behavior of the general flow parameter $t_{2k+1}$.\\\\
\begin{equation}
\begin{tabular}{ccc}\\\\
Flows  &\hspace{1cm} The higher n.l. terms  &\hspace{1cm}  Degree of n linearity \\ 
$t_{1}$       &   $u^{0}u'=u'$     &  $0$               \\ \\
$t_{3}$       &   $\frac{3}{2}uu'$       &\hspace{3cm}  $1$ \hspace{0,5cm} (quadratic)               \\ \\
$t_{5}$       &   $\frac{15}{2^{3}}u^{2}u'$       & \hspace{2,05cm}  $2$ \hspace{0,5cm}  (cubic)               \\ \\
$t_{7}$       &   $\frac{35}{2^{4}}u^{3}u'$       &  $3$               \\ \\
$t_{9}$       &   $\frac{315}{2^{7}}u^{4}u'$       &  $4$               \\\\
...           &...                 &...                     \\\\
$t_{2k+1}$    &  \hspace{0,5cm}  $ \eta {(2k+1)(2k-1)}u^{k}u'$   & \hspace{1cm} $(k)$, 

\end{tabular}
\end{equation}
where $\eta$ is an arbitrary constant.\\\\
This result shows among others that the $\theta$-evolution equations (95) exhibit at most a nonlinearity of degree $(k)$ associated to a term proportional to $(2k+1)(2k-1)u^{k}u'$. The particular case $k=0$ corresponds to linear wave equation.\\ \\
{\bf 3}. The contribution of non-commutativity to the Moyal KdV hierarchy shows a correspondence between the flows $t_{2k+1}$ and the non-commutativity parameters $\theta^{2(k-s)}, 0 \leq s \leq k$. Particularly, the higher term $\theta^{2(k)}$ is coupled to the $k-th$ prime derivative of $u_2$ namely $u^{(k)}$ while  the higher non linear term $\eta (2k+1)(2k-1)u^{k}u'$ is a $\theta$-independent object as its shown in eq.(95).\\\\
{\bf 4}. In analogy  with the classical case, once the non linear terms in the $\theta$-evolution equations are ignored, there will be no solitons in the KdV-hierarchy as the latter's are intimately related to non linearity [1].\\

\subsection{$sl_3$-Boussinesq Hierarchy}
The same analysis used in deriving the $sl_2$-KdV hierarchy is actually extended to build the $sl_3$-Boussinesq Moyal hierarchy. The latter is associated to the momentum Lax operator ${\mathcal L}_{3}= p^3 +u_2 \star p+ u_3$ whose $3-th$ root reads as 
\begin{equation}
\begin{array}{lcl}
{\mathcal L}^{\frac{1}{3}}&=& \Sigma_{i=-1} b_{i+1} \star p^{-i}\\\\
&=&{\Sigma}_{i=-1} a_{i+1}p^{-i} 
\end{array}
\end{equation}
in such way that ${\mathcal L}_{3}= {\mathcal L}^{\frac{1}{3}}\star {\mathcal L}^{\frac{1}{3}} \star {\mathcal L}^{\frac{1}{3}}$. 
Explicit computations lead to 
\begin{equation}
\begin{array}{lcl} 
b_0 &=& 1 \\\\
b_1 &=& 0 \\\\
b_2 &=&\frac{1}{3} u_{2} \\\\
b_3&=&\frac{1}{3}u_{3}-\frac{2}{3} \theta u_{2}^{'} \\\\

b_4&=&-\frac{1}{9}u_{2}^{2}-\frac{2}{3}\theta u_{3}^{'} +\frac{8}{9}\theta^{2} u_{2}^{''} \\\\

b_5&=&-\frac{2}{9} u_2 u_3 +\frac{8}{9}\theta u_2 u'_2 + \frac{8}{9} \theta^2  u''_3 -\frac{8}{9}\theta^3  u'''_3\\\\ 

b_6&=&\frac{1}{9}\{ \frac{5}{9}u_2^{3}-u_3^2 +2 \theta (4 u_2 u'_3 + 5 u'_2 u_3)-20\theta^2 (u_2 u''_2+(u'_2)^2)-8\theta^3 u'''_3 +\frac{16}{3}\theta^4u''''_2\} \\\\
b_7&=&\frac{1}{9} \{ \frac{5}{3}u_2^{2}u_3+10\theta(u_3 u'_3- u_2^2 u'_2) -\frac{20}{3} \theta^2 (5 u''_2 u_3 + 7 u'_2 u'_3+u_2 u'''_3)\\\\
& & -40\theta^3 (3u'_2 u''_2+u_2 u'''_2) + \frac{16}{3}\theta^4 u''''_3)\\ \\
b_{8}&=&{\frac{5}{27}} (u_{{2}}u_{{3}}^{2}-{\frac{2}{9}} u_{{2}}^{4}) -
\frac{10}{9}\theta(u_{2}^{2}u'_{3}-\frac{7}{3}u'_{2}u_{2}u_{3})
+\frac{20}{81}\theta ^{2} ( 12u_{3}^{2\prime }+31u_{2}u_{2}^{2\prime
}\\\\
& & +17u_{2}^{2}u''_{2}-15u''_{3}u_{3})+\frac{40}{27}\theta ^{3} ( 10u_{3}^{\prime \prime }u_{2}^{\prime
}+13u_{2}^{\prime \prime }u_{3}^{\prime }+7u_{3}u_{2}^{\prime \prime \prime
}+3u_{3}^{\prime \prime }u_{2})  \\\\
& & +\frac{80}{81}\theta ^{4} ( 8u_{2}^{4}u_{2}+23u_{2}^{2\prime
^{2}}+32u_{2}^{\prime }u_{2}^{\prime \prime \prime }) +{\frac{64}{81}}%
 {\theta }^{6}u_{{2}}^{(6)}\\\\
\end{array}
\end{equation}
Similarly, one can easily determine the coefficients $a_{i+1}$ which are also expressed as functions of $b_{i+1}$ and their derivatives. This result is summarized in the expression of ${\mathcal L}^{\frac{1}{3}}$ namely
\begin{equation}
\begin{array}{lcl}
{\mathcal L}^{\frac{1}{3}}= p & + & \frac{1}{3}u_{2}p^{-1} \\\\ 
 &+& \frac{1}{3}\{u_{3}-\theta u_{2}^{\prime }\}p^{-2} \\ \\
& - & \frac{1}{9}\{u_{2}^{2}+\theta ^{2}u_{2}^{\prime \prime }\}p^{-3} \\\\ 
& + & \frac{1}{9}\{-2u_{2}u_{3}+2\theta u_{2}^{\prime }u_{2}-\theta
^{2}u_{3}^{\prime \prime }+\theta ^{3}u_{2}^{\prime \prime \prime} \} p^{-4}
\\ \\
& + & \frac{1}{9} \{ \frac{1}{3} {\theta }^{4}u_{{2}}^{(4)}+2 \theta
u_{{2}}^{\prime }u_{{3}}- u_{{3}}^{2}+{\frac{5}{9}} u_{{2}}^{3} \} .{p}%
^{-5}   \\ \\
& + & \frac{1}{27} \{ {5} u_{{2}}^{2}u_{{3}}-{5} \theta  u_{{2}%
}^{2}u_{{2}}^{\prime }+{10} {\theta }^{2} ( u_{{2}}^{\prime }u_{{3}%
}^{\prime }-u_{{2}}^{\prime \prime }u_{{3}} ) + {\theta }^{4}u_{{3}%
}^{(4)}- {\theta }^{5}u_{{2}}^{(5)} \} .{p}^{-6} \\ \\
& + &\frac{1}{27} {\big \{ }
\frac{5}{9}u_{2} ( 9u_{3}^{2}-2u_{2}^{3} ) -{10} \theta  u_{{2}%
}^{\prime }u_{{2}}u_{{3}}+\frac{5}{3}\theta ^{2} ( 6u_{3}^{'2}-6u_{3}^{\prime \prime }u_{3}+5u_{2}^{2}u_{2}^{\prime \prime
}-2u_{2}u_{2}^{2\prime } )  \\ \\
&-&10\theta ^{3} ( -u_{3}^{\prime \prime }u_{2}^{\prime
}-u_{3}u'''_{2}+2u''_{2}u_{3}^{\prime
} ) -\frac{10}{3}\theta ^{4}( u_{2}^{(4)}u_{2}+4u_{2}^{\prime
}u'''_{2}-5u_{2}^{\prime \prime ^{2}} ) -{\frac{1}{%
3}} {\theta }^{6}u_{{2}}^{(6)}{\big \}}{p}^{-7}\\\\
&+&...
\end{array}
\end{equation}
Furthermore, using the Moyal $sl_3$-Lax evolution equations 
\begin{equation}
\frac{\partial {\mathcal L}}{\partial t_{k}}=\{({\mathcal L}^{\frac{k}{3}})_{+} ,{\mathcal L}\}_{\theta},
\end{equation}
that we compute explicitly for $k=1,2,4$ we obtain
\begin{equation}
\begin{array}{lcl}
\frac{\partial {\mathcal L}}{\partial t_{1}}&=&u'_{2}p+u'_{3}-{\theta}u''_{2}\\\\
\frac{\partial {\mathcal L}}{\partial t_{2}}&=&2\{u'_{3}-\theta u''_{2}\}p
-\frac{2}{3}\{u_{2}u'_{2}+{\theta}^{2}u'''_{2}\}\\\\
\frac{\partial {\mathcal L}}{\partial t_{4}}&=&\frac{4}{3}\{(u_{2}u_{3})'-\theta( u''_{2}u_2+u_{2}^{'2})+2{\theta}^{2}u'''_{3}- 2{\theta}^{3}u^{(4)}_{2}\}p\\\\
& & +\frac{4}{3}\{u_{3}u'_{3}-\frac{1}{3}u_{2}^{2}u'_{2}-\theta(u'_{2}u'_3+u''_{2}u_{3})- \theta^{2}(u'_{2}u''_2+u_{2}u'''_2)-\frac{2}{3}\theta^{4}u^{(5)}_2        \}\\\\
\end{array}
\end{equation} 
Identifying both sides of the previous equations, one obtain the following first leading evolution equations 
\begin{equation}
\begin{array}{lcl}
\frac{\partial}{\partial t_{1}}{u_2}&=&{u'_2}\\\\
\frac{\partial}{\partial t_{1}}{u_3}&=&{u'_3}\\\\\\
\frac{\partial}{\partial t_{2}}{u_2}&=&2{u'_3}-2\theta u''_2\\\\
\frac{\partial}{\partial t_{2}}{u_3}&=&-\frac{2}{3}u_{2}u'_{2}-\frac{8}{3}\theta^{2}u'''_{2}+2\theta u''_3\\ \\ \\  
\frac{\partial}{\partial t_{4}}{u_2}&=&\frac{4}{3}\{(u_{2}u_{3})'-\theta( u''_{2}u_2+u_{2}^{'2})+2{\theta}^{2}u'''_{3}- 2{\theta}^{3}u^{(4)}_{2}\}\\\\
\frac{\partial}{\partial t_{4}}{(u_3-\theta u'_{2})}&=& \frac{4}{3}\{u_{3}u'_{3}-\frac{1}{3}u_{2}^{2}u'_{2}-\theta(u'_{2}u'_3+u''_{2}u_{3})- \theta^{2}(u'_{2}u''_2+u_{2}u'''_2)-\frac{2}{3}\theta^{4}u^{(5)}_2    .  
\end{array}
\end{equation} 
These equations define what we call the Moyal $sl_3$ Boussinesq hierarchy. The first two equations are simply linear $\theta$-independent wave equations fixing the dimension of the first flow parameter $t_1$ to be $[t_{1}]=-1$. \\\\
The non trivial flow of this hierarchy starts really from the second couple of equations associated to $t_2$. We will discuss in the next section, how its important to deal with the basis of primary conformal fields $v_k$ instead of the old basis $u_k$. Anticipating this result, one can write the previous couple of equations in term of the spin $3$ primary field $v_3=u_3-\theta u'_2$ as follows
\begin{equation}
\begin{array}{lcl}
 \frac{\partial}{\partial t_{2}}{u_2}&=&2{v'_3}\\\\
\frac{\partial}{\partial t_{2}}{v_3}&=&-\frac{2}{3}\{u_{2}u'_{2}+\theta^{2}u'''_{2}\}
\end{array}
\end{equation} 
This couple of equations define the $\theta$-extended Boussinessq equation. Its second-order form is obtained by differentiating the first equation in eq.(103) with respect to $t_2$ and then using the second equation. We find
\begin{equation}
\frac{\partial^{2}}{\partial t^{2}_{2}} u_2=-\frac{4}{3}(u_{2}u'_{2} + {\theta}^{2}u_{2}^{(3)})',
\end{equation} 
Equivalently one may write 
\begin{equation}
\left(\matrix{u_2\cr v_3 \cr}\right)_{_{t_2}}=-\frac{2}{3}\left(\matrix {-3v'_2\cr {u_{2}u'_{2}+\theta^{2}u'''_2} \cr}\right)
\end{equation}
Recall that the classical Boussinesq equation is associated to the $sl_3$-Lax differential operator 
\begin{equation}
{\mathcal L}_{3}=\partial^{3}+2u\partial+v_3 
\end{equation}
with $v_3= u_3-\frac{1}{2}u'_2$ defining the spin-$3$ primary field.
This equation which takes the following form  
\begin{equation}
u_{tt}=-(auu' + bu^{(3)})',
\end{equation} 
where $a, b $ are arbitrary constants, arises in several physical applications. Initially, it was derived to describe propagation of long waves in shallow water [27]. This equation plays also a central role in $2d$ conformal field theories via its Gelfand-Dickey second Hamiltonian structure associated to the Zamolodchikov $w_3$ non linear algebra [8].\\\\ 
Actually all these important results are extended to the non-commutative $\theta$-deformed case [21-25].
Similarly the third couple of equations (102) can be equivalently written as
\begin{equation}
\begin{array}{lcl}
\frac{\partial}{\partial t_{4}}{u_2}&=&\frac{4}{3}(u_{2}v_{3} + 2\theta^{2}v''_{3} )'\\\\
\frac{\partial}{\partial t_{4}}{v_3}&=& \frac{4}{3}\{v_{3}v'_{3}-\theta^{2}u_{2}u'''_{2}-\frac{1}{3}(u^{2}_{2}u'_{2}+2\theta^{4}u_{2}^{(5)})\}
\end{array}
\end{equation} 
To close this section note that other flows equations associated to $(sl_2)$-KdV and $(sl_3)$-Boussinesq hierarchies can be also derived once some lengthly and hard computations are performed. One can also generalize the obtained results by considering other $sl_n$ integrable hierarchies with $n>3$.

\section{Dressing gauge group, covariantization of  $sl_n-\widehat \Sigma_{n}^{(0,n)}$ Lax operators and primarity condition}

The principal focus of this section is to present an alternative way to discuss conformal primary fields in the Das-Popowicz Moyal momentum algebra. Recall that the standard method for constructing primary fields $w_k$, is based on the well known covariantization method of Lax differential operators of Di-Francesco-Itzykson-Zuber (DIZ)[15].\\\\
This method has leads to build in a successful way all the $w_{k}$ conformal primary fields as functions of the old ones $u_k, k=2,3,4,...$ by covariantizing the corresponding Lax operators. Actually this method was extended and applied to the Moyal momentum case by the authors of [24] and has leads to build in a systematic way the $\theta$-deformation of the classical $w_k$-algebra. \\\\
Using some properties that we presented in an unpublished note [26], we will next show how one can interpret the DIZ-covariantization of Moyal momentum Lax operators presented in [24], as being a dressing gauge transformation in the Das-Popowicz Mm algebra. The idea consists in searching for the dressing gauge symmetry which ensures the primarity condition of the $w_k$-currents. This request makes strong constraints on the dressing gauge parameters and provides a geometrical interpretation of the primarity condition.\\\\ 
Note by the way that we will use two kinds of dressing operators: $K[a, u,V]$ and $K[a, u, w]$ corresponding to the dressing groups ensuring the transition from the ordinary basis $\{u_k\}$ to the intermediate basis $\{V_k\}$ and the primary basis $\{w_k\}$ respectively.

\subsection{The dressing gauge group of  $sl_n-\widehat \Sigma_{n}^{(0,n)}$}
Let us first start by defining the dressing gauge symmetry group. Elements of this  group (called also the Volterra gauge group)[26] are given by the Lorentz scalar momentum operators
\begin{equation}
K[a]=1+\sum_{j}a_{j} \star p^{-j}
\end{equation}
where $a_{j}\equiv a_{j}(x,t)$ are arbitrary analytic functions of conformal spin $j\in Z$. This momentum operator belonging to ${\widehat \Sigma}_{0}^{(-\infty, 0)}$ can be also written in ${\Sigma}_{0}^{(-\infty, 0)}$ as
\begin{equation}
\begin{array}{lcl}
K[a]&=&1+a_{1}p^{-1}+(a_2+\theta a'_1)p^{-2}+(a_3+\theta^{2}a_1+2\theta a'_2)p^{-3}+...\\\\
&=&1+\sum_{m=1}^{\infty}\sum_{s=0}^{m-1}{\theta^{s}}c^{s}_{m-1}a^{(s)}_{m-s}p^{-m}
\end{array}
\end{equation}
The functions $a_i$, to which we shall refer hereafter to as the dressing gauge parameters, can be expressed in terms of the residue operation as follows 
\begin{equation}
a_{j}={\widehat Res}\{K[a]\star p^{j-1}\}, j=1,2,3,...
\end{equation}
with 
\begin{equation}
\begin{array}{lcl}
a_{1}={\widehat Res}K[a]\\
a_{2}={\widehat Res}(K[a]\star p)\\
a_{3}={\widehat Res}(K[a]\star p^2)\\
.\\
.\\
.
\end{array}
\end{equation}
Next, we show how one can see the dressing group as a gauge symmetry of the space $sl_{n}-\widehat \Sigma_{n}^{(0,n)}$. The idea consists in considering the following mapping
\begin{equation}
{{\mathcal L}_{n}}(u)=\sum _{i=0}^{n}u_{n-i}\star p^i \longrightarrow  {{\mathcal L}_{n}}(V)=\sum _{i=0}^{n}V_{n-i}\star p^i,
\end{equation}
where ${\mathcal L}_{n}(u)$ is an $sl_{n}-\widehat \Sigma_{n}^{(0,n)}$ Lax operator, and searching for the explicit form of the dressing operators $K[a,u, V]\equiv K[a]$ for which the operators ${\mathcal L}_{n}(V)$ given by
\begin{equation}
{{\mathcal L}_{n}}(V)=K^{-1}[a]\star{{\mathcal L}_{n}}(u)\star K[a],
\end{equation}
are also $sl_{n}-\widehat \Sigma_{n}^{(0,n)}$ Lax operators.\\\\
To be more precise, let's consider for simplicity the following first non trivial example 
\begin{equation}
\begin{array}{lcl}
{\mathcal L}_3&=&p^3+u_2\star p+u_3\\
&=&p^3+u_2.p+(u_3-\theta u'_2).
\end{array}
\end{equation}
Using eq.(114) and performing lengthy computations, we present here bellow, the first leading coefficient-functions of $p^i$ , up to $p^-3$,\\
\begin{equation}
\begin{array}{lcl}
0&=&{{\mathcal L}_{3}}(u)\star K[a]-K[a]\star {{\mathcal L}_{3}}(V)\\ \\
&=&\{u_2-V_2\}p\\ \\
&+ &\{6\theta a'_2+u_3-V_3\}p^0\\ \\
&+ &\{12{\theta}^2 a''_2+6\theta a'_3\}p^{-1}\\ \\
&+ &\{18{\theta}^2 a''_3+6\theta a'_4+20{\theta}^3 a'''_2+u_3 a_2 + 2\theta u_2 a'_2+4\theta a_2 u'_2-a_2 V_3\}p^{-2}\\ \\
&+ &\{12\theta^{2} u'_{{2}}a'_{{2}}+38{\theta}^3 a'''_3+ 24{\theta}^2 a''_4- a_3 V_3 +28{\theta}^4 a^{(4)}_2+ 6{\theta} a'_5+u_3 a_3+ 6{\theta} u'_2 a_3\\ \\
&- & 4{\theta}^2 u''_2 a_2+2{\theta} V'_3 a_2+4{\theta}^2 a''_2 u_2+2{\theta} u_2 a'_3-2{\theta} a'_2 V_3+2{\theta} u'_3 a_2+2{\theta} a'_2 u_3\}p^{-3}\\ \\
& &. . .
\end{array}
\end{equation}\\
from which we can derive the constraint equations $a_k=a_{k}\{V^{(l)}, u^{(l)}\}, l=0,1,2,...$. with $V^{(1)}=V'$ is the prime derivative.\\\\
The next step is to solve these equations, which means to find the explicit form of the dressing gauge parameters $a_i$. This job seems to be tedious as we ignore for the moment how one can express the new fields $V_i$ in terms of the old ones $u_i$.\\
Simplifying eqs.(116), with the choice $(a_1=0)$, we obtain $V_2=u_2$ and 
\begin{equation}
\begin{array}{lcl}
a'_2&=&-\frac{1}{6\theta}\{u_3-V_3\}\\\\
a_3&=&\frac{1}{3}\{u_3-V_3\}\\\\
a'_4&=&\frac{1}{6\theta}\{a_2 V_3-u_3 a_2 -2\theta (u_2 a'_2+2 a_2 u'_2)-18{\theta}^2 a''_3-20{\theta}^3 a'''_2\}\\\\
a'_5 &=& -\frac{1}{6\theta}\{ 12\theta^{2} u'_{{2}}a'_{{2}}+38{\theta}^3 a'''_3+ 24{\theta}^2 a''_4- a_3 V_3 +28{\theta}^4 a^{(4)}_2+ 6{\theta} a'_5+u_3 a_3+ 6{\theta} u'_2 a_3\\ \\
& & - 4{\theta}^2 u''_2 a_2+2{\theta} V'_3 a_2+4{\theta}^2 a''_2 u_2+2{\theta} u_2 a'_3-2{\theta} a'_2 V_3+2{\theta} u'_3 a_2+2{\theta} a'_2 u_3 \}\\\\
...
\end{array}
\end{equation}
Using dimensional arguments, we can set 
\begin{equation}
a_2=\alpha u_2,
\end{equation}
where $\alpha$ is a coefficient number that we can explicitly fix. With this value of $a_2$ one can then split the set of equations (117) into two kinds of equations namely
\begin{equation}
\begin{array}{lcl}
V_2&=&u_2\\
V_3&=&u_3+6\theta\alpha u'_2
\end{array}
\end{equation}
giving the expressions of the new fields $(V_2 , V_3)$ in terms of the old ones $(u_2 , u_3)$, and
\begin{equation}
\begin{array}{lcl}
a_3&=&-2\theta \alpha u'_2\\\\
a_4&=&\frac{\alpha^2 - \alpha}{2}u^{2}_{2}+\frac{8}{3}\theta^{2}\alpha u^{''}_{2}\\\\
a'_5 &=& \frac{8}{3}\theta^{3}\alpha u_{2}^{(4)}-4\theta \alpha(\alpha - 1){u^{'}_2}^{2}-\frac{2}{3}\alpha u_{2}u'_{3}-4\theta \alpha(\frac{3}{2}\alpha - \frac{7}{6})u_{2}{u^{''}_2} \\ \\
...
\end{array}
\end{equation}
Later on we will discuss  the particular situation where the fields $V_i$ are considered as primary fields a fact which makes the solving of the dressing gauge parameters much more easiest and also interesting. In fact, as we are working in $sl_{3}-{\widehat \Sigma}_{3}^{(0,3)}$ and since $V_3-u_3=6\alpha\theta u'_2$, we can easily show that a natural realization of $\alpha$ is given by $\alpha =-\frac{1}{6}$. With this fixed value, the previous equations reduce simply to
\begin{equation}
\begin{array}{lcl}
a_2&=&\frac{-1}{6}u_2\\\\
a_3&=&\frac{1}{3}\theta u'_2\\\\
a_4&=&-\frac{4}{9}{\theta}^{2}u''_2+\frac{7}{72}u_{2}^{2}\\\\
a'_5&=&\frac{4}{9}{\theta}^{3}u^{(4)}_2 - \theta(\frac{17}{18}u''_{2}u_{2} + \frac{7}{9}{u'_2}^{2})+\frac{1}{9}u'_{3}u_2\\

\end{array}
\end{equation}\\
The derived eqs.(121) correspond to the parameters of the dressing gauge group $K[a, u, V]$ ensuring the passage from the old basis $\{u_2, u_3\}$ to the new primary conformal basis $\{V_2, V_3\}$ where $V_3$ is the primary field of conformal spin 3 which reads by virtue of eq.(119) as $V_3=u_3-\theta u'_2$.\\\\
We remark through this example that there exist two kinds of parameters. The first kind stands for the  gauge dressing parameters $a_i$, $1\leq i \leq 4$ which are well defined  and the second kind contains the parameters $a_i, i>4$ which are not well defined as shown for example for $a_5$.\\\\ 
The undefined parameters are to be considered as extra gauge dressing parameters. In fact the knowledge of the first dressing parameters $a_i$, $1\leq i \leq 4$ is enough to ensure the primarity condition for the fields $\{u_2, u_3\}$ and then to build the dressing gauge group associated to $sl_3-{\widehat \Sigma}^{(0,3)}_{3}$\\

The second example that we give is the $sl_4$ case given by
\begin{equation}
\begin{array}{lcl}
{\mathcal L}_4&=&p^4+u_{2}\star p^2+u_{3}\star p+u_4\\
&=&p^4+u_{2}p^{2}+(u_3-2\theta u'_2)p+u_4-\theta u'_3+\theta^2 u''_2.
\end{array}
\end{equation}
Explicit computations give 
\begin{equation}
\begin{array}{lcl}
0&=&{{\mathcal L}_{4}}(u)\star K[a]-K[a]\star {{\mathcal L}_{4}}(V)\\ \\
&+& (u_{3}-V_{3}+8\theta a'_{2}) p \\ \\
&+ & ( u_{4}-V_{4}+\theta ( V'_{3}-u'_{3}+8a'_{3}) +16\theta^{2}a''_{2})\\\\
& +& \{ a_{2}(u_{3}-V_{3})+4\theta ( 2a_{4}+a_{2}u_{2}+6\theta a'_{3}+8{\theta}^{2}a''_{2})'\}p^{-1}\\\\
& +& \{ a_{2}(u_{4}-V_{4})+ a_{3}(u_{3}-V_{3})+\theta ( 8a'_5+6a_{3}u'_2++a_{2}u'_{3}+4u_{2}a'_{3}+3a_{2}V'_3+3u_{3}a'_2-a'_{2}V_3)\\\\
&+&8\theta^{2}(4a''_4+u'_{2}a'_2-a_{2}u''_2+u_{2}a''_2)+56\theta^{3}a'''_3+48\theta^{4}a^{(4)}_{2}\}p^{-2}
\\\\
&...&
\end{array}
\end{equation}
once we set $a_{1}=0$ corresponding to $u_{2}=V_{2}$. \\\\
Furthermore, setting once again $a_{2}=\alpha u_{2}$, we obtain 
\begin{equation}
u_{3}-V_{3}=-8\theta \alpha u'_{2}
\end{equation}
and the following first leading dressing gauge parameters
\begin{equation}
\begin{array}{lcl}
a'_{3}&=&-3\alpha \theta  ( u''_{2})+\frac{1}{8\theta } ( V_{4}-u_{4})\\\\ 
a_{4}&=&\frac{1}{2}\alpha  ( \alpha -1) u_{2}^{2}+\frac{3}{8} (u_{4}-V_{4})+5\theta^{2}\alpha u''_{2}\\\\
a'_{5}&=&( \alpha -\frac{3}{4})a_{3}u'_{2}-\frac{1}{4}\alpha(u'_{2}u_{3}+2u_{2}u'_{3})\\\\
&+&\theta \{\frac{5}{8}( V_{4}-u_{4})''+3\alpha(\alpha -1)u_{2}^{'2}+\alpha( 
\frac{11}{2}-7\alpha ) u_{2}u''_{2}\}\\\\
&-&5\theta ^{3}\alpha u_{2}^{(4)}+\frac{1}{8\theta } ( \alpha -\frac{1%
}{2}) u_{2} ( V_{4}-u_{4}) \\\\
&...&

\end{array}
\end{equation}

Actually, the techniques developed for the previous two examples $(n=3,4)$, are generalizable to $sl_n-{\widehat \Sigma}_{n}^{(0,n)}, n\geq 5$ and the same induced algebraic properties are shown to apply also for higher order conformal spin gauge parameters $a_i$, $i>5$. We have for arbitrary $n$\\

\begin{equation}
{\mathcal L}_n=p^n+u_{2}\star p^{n-2}+u_{3}\star p^{n-3}+u_4\star p^{n-4}+...+u_{n-1}\star p+u_{n}\\
\end{equation}
and
\begin{equation}
\begin{array}{lcl}
0&=&{{\mathcal L}_{n}}(u)\star K[a]-K[a]\star {{\mathcal L}_{n}}(V)\\ \\
&=& (u_{3}-V_{3}+2n\theta a'_{2}) p^{n-3} \\ \\
&+ & (u_4-V_4+(n-3)\theta(V'_3-u'_3)+2n \theta a'_3 +4n \theta^2 a''_2)p^{n-4}\\\\
& & \vdots 
\end{array}
\end{equation}
These generalized expressions show among others that the parameter of the gauge dressing group are given by
\begin{equation}
\begin{array}{lcl}
a_{1}&=&0\\\\
a_{2}&=&\frac{2-n}{2n}u_2\\\\
a'_3&=&\frac{7 \theta (n-2)(n-3)}{10n} u''_2-
\frac{n-3}{2n} u'_3 - \frac{(n-2)(n-3)(5n+7)}{20 \theta n (n^3-n)} u_2^2
 -2 \theta a''_2\\
& & \vdots
\end{array}
\end{equation}
Setting $n=3,4$ associated to ${\mathcal L}_3$ and ${\mathcal L}_4$ we recover the values of $a_2=-\frac{1}{6}u_2,-\frac{1}{4}u_2$ respectively.\\\\
To summarize, we can then  construct a gauge symmetry group of the Das-Popowicz momentum subspaces $sl_n-{\widehat \Sigma}_{n}^{(0,n)}$ which is nothing but the dressing group whose elements are the Lorentz scalar dressing momentum operators $K[a]$. Requiring the invariance of $sl_n$ momentum Lax operators under the action of this dressing group makes strong constraint on the gauge parameters $a_i$ as its explicitly shown in the $sl_3$ and $sl_4$ described examples. \\\\
These techniques are also important as they allow us to interpret the mapping 
\begin{equation}
\{u_i\}\stackrel{K[a]}{ \longrightarrow}\{V_i\}
\end{equation}
 between two different basis of conformal fields as being a gauge choice on some orbit that we call the dressing gauge orbit. Elements of this orbit are well defined gauge dressing groups $\{K_{i}[a]\}$ such that each position on this orbit is characterized by a fixed dressing operator $K_{i}[a]$ which fixes in turn the gauge group.\\\\ 
The basis of conformal primary fields $\{w_k\}$, that we will discuss later, may be interpreted as a choice of a position on the gauge orbit. This position corresponds to a particular dressing gauge operator $K_{p}[a]\equiv K[u, u, w]$ whose parameter are the ones ensuring the transition from the ordinary basis $\{u_i\}$ to the primary conformal one $\{w_i\}$.\\\\

We represent this orbit as follows:\\\\

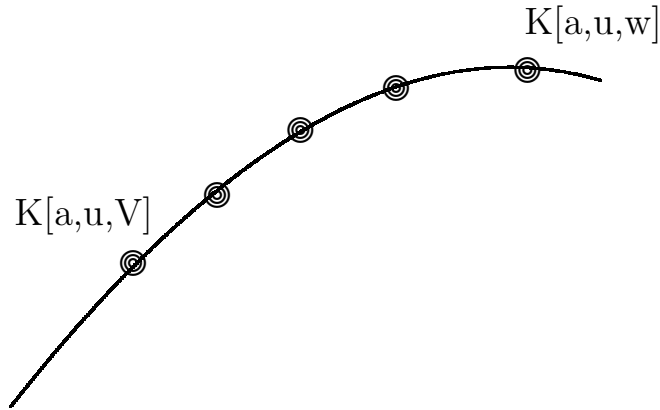
\begin{center}
\setlength{\unitlength}{3947sp}%
\begingroup\makeatletter\ifx\SetFigFont\undefined%
\gdef\SetFigFont#1#2#3#4#5{%
  \reset@font\fontsize{#1}{#2pt}
  \fontfamily{#3}\fontseries{#4}\fontshape{#5}
  \selectfont}
\fi\endgroup%

\begin{picture}(3915,1920)(901,-1951)
\thicklines

\bezier{1000}(800,-2551)(2800,0)(4500,-500)

\put(3901,-211){\makebox(0,0)[lb]{\smash{\SetFigFont{14}{16.8}{\rmdefault}{\mddefault}{\updefault}{K[a,u,w]}}}
}
\put(700,-1411){\makebox(0,0)[lb]{\smash{\SetFigFont{14}{16.8}{\rmdefault}{\mddefault}{\updefault}{K[a,u,V]}}}
}

\put(1576,-1650){\circle{150}}
\put(1576,-1650){\circle{100}}
\put(1576,-1650){\circle{50}}

\put(2101,-1220){\circle{150}}
\put(2101,-1220){\circle{100}}
\put(2101,-1220){\circle{50}}

\put(2626,-811){\circle{150}}
\put(2626,-811){\circle{100}}
\put(2626,-811){\circle{50}}

\put(3226,-550){\circle{150}}
\put(3226,-550){\circle{100}}
\put(3226,-550){\circle{50}}

\put(4051,-436){\circle{150}}
\put(4051,-436){\circle{100}}
\put(4051,-436){\circle{50}}
\end{picture}
\end{center}

\vspace{2cm}
\subsection {The natural ``primarity'' realization}
Having discussed the structure of the gauge symmetry group of the Das-Popowicz momentum subspaces $sl_n-{\widehat \Sigma}^{(0,n)}_{n}$, we will now show how one can naturally realize the  ${v}_n$-currents just by exploiting the definition of the star product. The notation $\{v_n\}$ is introduced to refer to the natural ``primary'' fields basis distinguished from the $\{w_k\}$ primary basis obtained from the $\theta$-extended DIZ covariantization mechanism [24].\\\\
In fact consider the following momentum $sl_n$-Lax operator 
\begin{equation}
{{\mathcal L}_{n}}(u)=\sum _{i=0}^{n}u_{n-i}\star p^i,
\end{equation}
for which we set $u_{0}=1$ and $u_{1}=0$. As discussed previously, these $sl_n$-Lax operators are simply elements of the coset space ${\widehat \Sigma}_{n}^{(0,n)}/{\widehat \Sigma}_{n}^{(1,1)}$ that we denote as  $sl_{n}$-${\widehat \Sigma}_{n}^{(0,n)}$. One easily check that the $sl_n$-Lax operators can be written in the space $sl_{n}$-$\Sigma_{n}^{(0,n)}$ as follows
\begin{equation}
{{\mathcal L}_{n}}(v)=\sum _{i=0}^{n}v_{n-i}p^i
\end{equation}
where the $v_i$ are conformal fields depending on the ordinary ones $\{u_i\}$ and their  derivatives as well as in the $\theta$ parameter. We easily show that the set of equations for the $v_i$-fields is explicitly given by
\begin{equation}
\begin{array}{lcl}
v_{2}=u_2\\
v_{3}=u_{3}-(n-2)\theta u'_{2}\\
v_{4}=u_{4}-(n-3)\theta u'_{3}+\frac{{\theta^2}(n-2)(n-3)}{2}u''_2\\
v_{5}=u_{5}-(n-4)\theta u'_{4}+\frac{{\theta^2}(n-3)(n-4)}{2}u''_3-\frac{{\theta^3}(n-2)(n-3)(n-4)}{6}u'''_{2}\\
.\\
.\\
.\\
v_{n}=u_{n}-\theta u'_{n-1}+ {\theta^2} u''_{n-2}-{\theta^3} u'''_{n-3}+...+(-1)^{n}{\theta}^{n-2}u_{2}^{(n-2)}.
\end{array}
\end{equation}
These expressions show the possibility to realize, in a  natural way, the conformal fields $v_i$ in terms of the ordinary ones $\{u_i\}$ and their derivatives just by looking at the  momentum Lax operators of $sl_{n}$-${\widehat \Sigma}_{n}^{(0,n)}$ as being also momentum Lax operators of $sl_{n}$-${\Sigma}_{n}^{(0,n)}$. Note also that the first relations
\begin{equation}
\begin{array}{lcl}
v_{2}=u_2\\
v_{3}=u_{3}-(n-2)\theta u'_{2},
\end{array}
\end{equation}
coincide exactly with the result obtained by using the DIZ covariantization of Moyal momentum Lax operators as done in [24]. The $v_{2}=u_2$ is nothing but the the analogue of the spin 2 conformal current of 2-dimensional CFT theories and which does not transform as a primary field and $v_3$ is the primary spin 3 current.\\\\
The major difference between the two approaches is that in the natural realization,  we do not require anything for the fields $v_i$ wile in the Moyal DIZ covariantization a condition for these new fields to be primary is required.\\\\
Nevertheless, forgetting about the non-linear terms in the primary basis realization in [24], one can easily observe the striking resemblance with the natural realization eqs.(132). To illustrate this point, let's consider the following result established by the authors of [24] for the primary $w_k$-fields
\begin{equation}
\begin{array}{lcl}
w_{3}&=&u_{3}-(n-2)\theta u'_{2}\\ \\
w_{4}&=&u_{4}-\frac{(n-2)(n-3)(5n+7)}{10(n^3-n)}u^{2}_2-(n-3)\theta u'_{3}+\frac{{2\theta^2}(n-2)(n-3)}{5}u''_{2} \\ \\
w_{5}&=&u_{5}-(n-4)\theta u'_{4}+\frac{(n-3)(n-4)(7n+13)}{7(n^3-n)}[\theta(n-2)u_{2}u'_{2}-u_{2}u_{3}]\\\\
&+&\frac{{3\theta^2}(n-3)(n-4)}{7}u''_3-\frac{{2\theta^3}(n-2)(n-3)(n-4)}{21}u'''_{2}\\\\
&.....&
\end{array}
\end{equation}
 
The linear truncation of eqs.(134) leads to the same global form of the natural realization eqs.(132). 
One can then conclude that the star product framework provides a natural realization of the primary conformal field $v_3$ without any condition. For higher spin fields, $v_i$, $i=4,5,...$, the natural realization can be seen as a linear truncation of the DIZ covariantization a la Moyal [24].\\\\
Now its important to look for the dressing gauge group ensuring the transition to the natural primary basis $\{v_i\}$ eqs.(132). Straightforward computations based on the previous analysis lead to the following result  corresponding the $sl_4$-case.\footnote{The different mapping of $sl_3$-momentum Lax operators ${\mathcal L}_{3}(u) \stackrel{K[a]}{ \longrightarrow} {\mathcal L}_{3}(v)$, ${\mathcal L}_{3}(V)$ or ${\mathcal L}_{3}(w)$ correspond to the same realization of the spin-$3$ field, namely $v_3=V_3=w_3=u_3-\theta u'_{2}$.} 
\\\\
Using the natural realization eqs.(132) we are now in position to solve the derived expressions eqs.(123-125), which means to find the form of the associated gauge dressing group. Substituting eqs.(132) for $n=4$ into eqs.(123-125) one obtain $\alpha =-\frac{1}{4}$ and    
\begin{equation}
\begin{array}{lcl}
v_2-u_2&=&0\\\\
v_3-u_3&=& -2\theta u'_2\\\\
v_4-u_4&=& -\theta u'_3+\theta^{2}u''_2\\\\
\end{array}
\end{equation}
and
\begin{equation}
\begin{array}{lcl}
a_2&=&-\frac{1}{4}u_2\\\\
a_3&=& -\frac{1}{8}u_3+\frac{7}{8}\theta u'_2\\\\
a_4&=& \frac{5}{32}u_{2}^{2}+\frac{3}{8}\theta u'_3-\frac{13}{8}\theta^{2}u''_2\\\\
a'_5&=&\frac {15}{8}{\theta}^{3} u^{(4)}_{2} -\frac{5}{8}\theta^{2} u'''_{3}-\frac{1}{16}\theta(29u_{2}^{'2} +\frac {61}{2}u''_{2} u_{2})\\\\
&+&{\frac {7}{32}}u_{{2}}u'_{{3}}+\frac{3}{16}u_{{2}}u_{{3}}\\\\
&...&
\end{array}
\end{equation}
These parameters define the gauge dressing group that we are searching for. 

\subsection {The gauge group of primary fields}
The dressing gauge symmetry group is used once again to interpret the primarity condition of higher spin fields as being a gauge choice on the so called dressing gauge symmetry group. The idea consist in considering the dressing operators eq.(109) such that the following transformation
\begin{equation}
{{\mathcal L}_{n}}(u)=\sum _{i=0}^{n}u_{n-i}\star p^i \longrightarrow  {{\mathcal L}_{n}}(w)=\sum _{i=0}^{n}w_{n-i}\star p^i,
\end{equation}
with
\begin{equation}
{{\mathcal L}_{n}}(w)=K^{-1}[a]\star{{\mathcal L}_{n}}(u)\star K[a],
\end{equation}
corresponds to the passage from the ordinary basis $\{u_k\}$ to the associated conformal primary one $\{w_k\}$. The primarity condition of the fields $w_k$ eqs.(134), makes strong constraint on the dressing parameters $a_i$ whose knowledge is sufficient to determine completely the dressing gauge group $K[a, u, w]\equiv K[a]$.\\\\
To do so, let's start from the $n-th$ order $sl_n$-momentum Lax operators ${{\mathcal L}_{n}}(u)$ eq.(126) and consider  the following  relation
\begin{equation}
{{\mathcal L}_{n}}[\tilde u]=J^{-\frac{n+1}{2}}\star{{\mathcal L}_{n}}[u]\star J^{-\frac{n+1}{2}}.
\end{equation}
This formula, generalizing the DIZ standard one [15] to the non-commutative case, is the key step towards deriving the primarity condition as well as the conformal transformation law of the $u_i$ conformal fields under the diffeomorphism
\begin{equation}
x \longrightarrow {\tilde x}(x),             
\end{equation}
where $\tilde u$ is the transform of $u$ under the conformal change eq(140) and where the conjugate momenta of $\tilde x$ with respect to the Moyal bracket is given by
\begin{equation}
\tilde p = J^{-1}\star p
\end{equation}
such that $\{\tilde p, \tilde x\}_{\theta}=1$. The symbol $J=J(x)$ is the Jacobian of the transformation eq.(140) given by $J(x)\equiv \frac{\partial \tilde x}{\partial x}$\\\\
What we find in general is that the  $u$-currents do not obey the primarity condition. One needs to restore this primarity property a fact which puts all the higher spin currents into Virasoro representations. The procedure is that one may perform adequate change of variables in the space of the  $u$-fields as done by D-Francesco et al [15] in the classical framework and by M.H.Tu et al [24] in the Moyal deformation case.\\\\
Our principal next goal is to show that this procedure is nothing but the result of a dressing gauge transformation in the space $\Sigma_{k}^{(0,0)}$ of functions of conformal spin $k\in Z$.\\\\
To illustrate this result recall once again, that the dressing group acts on the algebra $sl_{n}-{\widehat \Sigma}_{n}^{(0,n)}$  through its adjoint representation. Using eq(138), one find that the $w_k$-currents may be expressed completely into the gauge parameters $a_k$, the fields $u_k$ and their $k$-th derivatives. \\\\
Next, using the fact that the $w_k$'s are primary fields, one can determine the corresponding dressing gauge group. To show how these things work, let's start by solving eq.(138) for the first non trivial example based on the $sl_4 -{\widehat \Sigma}^{(0,4)}_{4}$ Lax operator. 
\begin{equation}
\begin{array}{lcl}
{\mathcal L}_4&=&p^4+u_{2}\star p^2+u_{3}\star p+u_4\\
&=&p^4+u_{2}p^{2}+(u_3-2\theta u'_2)p+u_4-\theta u'_3+\theta^2 u''_2.
\end{array}
\end{equation}
Consider eq.(123) for the primary fields $w_k$
\begin{equation}
\begin{array}{lcl}
0&=&{{\mathcal L}_{4}}(u)\star K[a]-K[a]\star {{\mathcal L}_{4}}(w)\\ \\
&=& (u_{3}-w_{3}+8\theta a'_{2}) p \\ \\
&+ & ( u_{4}-w_{4}+\theta ( w'_{3}-u'_{3}+8a'_{3}) +16\theta^{2}a''_{2})\\\\
& +& \{ a_{2}(u_{3}-w_{3})+4\theta ( 2a_{4}+a_{2}u_{2}+6\theta a'_{3}+8{\theta}^{2}a''_{2})'\}p^{-1}\\\\
& +& \{ a_{2}(u_{4}-w_{4})+ a_{3}(u_{3}-w_{3})+\theta ( 8a'_5+6a_{3}u'_2++a_{2}u'_{3}+4u_{2}a'_{3}+3a_{2}w'_3+3u_{3}a'_2-a'_{2}w_3)\\\\
&+&8\theta^{2}(4a''_4+u'_{2}a'_2-a_{2}u''_2+u_{2}a''_2)+56\theta^{3}a'''_3+48\theta^{4}a^{(4)}_{2}\}p^{-2}
\\\\
&...&
\end{array}
\end{equation}
and using the result, eqs(134), found in [24] for the first primary fields namely
\begin{equation}
\begin{array}{lcl}
w_{3}&=&u_{3}-2\theta u'_{2}\\ \\
w_{4}&=&u_{4}-\frac{9}{100}u^{2}_2-\theta u'_{3}+\frac{{4\theta^{2}}}{5}u''_{2} \\ \\
\end{array}
\end{equation}
 
Then, substituting these expressions of the primary fields into eq(143) we obtain after solving the equations
\begin{equation}
\begin{array}{lcl}
a_2&=&-\frac{1}{4}u_2\\\\
a'_3&=&\frac {1}{800\theta}\{ 9u_{2}^{2}+100\theta u'_3-280{\theta}^{2}u_2\} \\\\
a_4&=&\frac {49}{400}u_{2}^{2}-\frac{3}{8}\theta u'_3+\frac{41}{20}{\theta}^{2}u''_2 \\\\
\vdots
\end{array}
\end{equation}
The general situation consists in considering, as previously, the $sl_n$-${\widehat \Sigma}_{n}^{(0,n)}$ Lax operators
\begin{equation}
{\mathcal L}_n=p^n+u_{2}\star p^{n-2}+u_{3}\star p^{n-3}+u_4\star p^{n-4}+...+u_{n-1}\star p+u_{n}\\
\end{equation}
satisfying the following gauge symmetry transformation, under the action of the dressing gauge group $K[a,w]$,
\begin{equation}
\begin{array}{lcl}
0&=&{{\mathcal L}_{n}}(u)\star K[a]-K[a]\star {{\mathcal L}_{n}}(w)\\ \\
&=& (u_{3}-w_{3}+2n\theta a'_{2}) p^{n-3} \\ \\
&+ & (u_4-w_4+(n-3)\theta(w'_3-u'_3)+2n \theta a'_3 +4n \theta^2 a''_2)p^{n-4}\\\\
& & \vdots 
\end{array}
\end{equation}
These generalized expressions are simplified by setting $w_2 = u_2$ which corresponds to the vanishing of the coefficient of $p^{n-2}$. They show among others that the parameter of the gauge dressing group associated to the primarity condition read as
\begin{equation}
\begin{array}{lcl}
a_{1}&=&0\\\\
a_{2}&=&\frac{2-n}{2n}u_2\\\\
a'_3&=&\frac{7 \theta (n-2)(n-3)}{10n} u''_2-
\frac{n-3}{2n} u'_3 - \frac{(n-2)(n-3)(5n+7)}{20 \theta n (n^3-n)} u_2^2
 -2 \theta a''_2\\
& & \vdots
\end{array}
\end{equation}
where we have used eqs.(134) giving the expressions of the primary fields $w_k$ in terms of the fields $u_k$.

\section{Concluding remarks}

Let us summarize the content of this work:  \\

We have presented a systematic study of so called Moyal momentum algebra (Mm algebra)[21] that we call the Das-Popowicz Mm algebra, denoted in our convention notation by ${\widehat \Sigma}_{\theta}$ [25]. This is the huge space of momentum Lax operators of arbitrary conformal spin $m$,$m\in \small Z$ and arbitrary higher and lowest degrees $(r,s)$ reading as
\begin{equation}
{\tilde{\mathcal  L}}_{m}^{(r,s)}(u)=\sum _{i=r}^{s}p^{i}\star u_{m-i}
\end{equation}
We study the algebraic properties of  ${\widehat \Sigma}_{\theta}$ and its underlying sub-algebras  ${\widehat \Sigma}_{m}^{(r,s)}$ and show that among all these spaces  only the subspace ${\widehat \Sigma}_{0}^{(-\infty, 1)}$ which defines a Lie algebra structure with respect to the Moyal bracket. \\\\
We define two kind of residue operations, exhibiting both a conformal spin equal to 1 and which act on two different spaces but with value on the ring ${\widehat \Sigma}^{(0,0)}$.\\\\
We introduce a degrees pairing product 
\begin{equation}
\left(.,.\right): {\widehat \Sigma}_{m}^{(r,s)} \star {\widehat \Sigma}_{n}^{(-s-1,-r-1)} \rightarrow  {\Sigma}_{m+n+1}^{(0,0)},
\end{equation}
showing that the spaces ${\widehat \Sigma}_{m}^{(r,s)}$ and  ${\widehat \Sigma }_{n}^{(-s-1 ,-r-1)}$ are $\widehat Res$-dual as do the subspaces ${\Sigma}_{m}^{(r,s)}$ and  ${\Sigma }_{n}^{(-s-1 ,-r-1)}$ with respect to the $Res$-operation [7].\\\\
We introduce also a combined scalar product $\left<\left<, \right>\right>$ which carries the right conformal spin quantum number namely $[\left<\left<, \right>\right>]=0$. This is very important towards building the $\theta$ analogue of the GD second Hamiltonian structure.\\\\
The particular sub-algebra $sl_{n}-{\widehat \Sigma}_{n}^{(0,n)}$ built out of the $sl_{n}$ momentum Lax operators ${\tilde{\mathcal  L}}_{n}^{(0,n)}(u)=\sum _{i=0}^{n}p^{i}\star u_{n-i}$, with $u_0=1$ and $u_1=0$, is applied to field theory building. Indeed, using the properties of this sub-algebra we were able to construct the $\theta$-Liouville conformal model 
\begin{equation}
\partial{\bar \partial}\phi=\frac{2}{\theta}e^{-\frac{1}{\theta}\phi}
\end{equation}
and its $sl_3$-Toda extension. 
\begin{equation}
\begin{array}{lcl}
\partial{\bar \partial}\phi_1&=&Ae^{-\frac{1}{2\theta}(\phi_1+\frac{1}{2}\phi_2)}\\\\
\partial{\bar \partial}\phi_2&=&Be^{-\frac{1}{2\theta}(\phi_1+2\phi_2)}
\end{array}
\end{equation}\\
We show also that the central charge, a la Feigin-Fuchs, associated to the spin-2 conformal current of the $\theta$-Liouville model is given by
\begin{equation}
c_{\theta}=(1+24\theta^{2})
\end{equation}\\
The results obtained for the Das-Popowicz Mm algebra, are applied to study some properties of $sl_2$-KdV and $sl_3$-Boussinesq integrable hierarchies. Our contributions to this study consist in extending the results found in [23] by increasing the order of computations a fact which leads us to discover more important properties as its explicitly shown in sec(4).\\\\ 
As an original result, we build the $\theta$-deformed $sl_3$-Boussinesq hierarchy and derive the associated $\theta$-flows. The second flow corresponds to the $\theta$-Boussinesq equation given by
 \begin{equation}
\left(\matrix{u_2\cr v_3 \cr}\right)_{_{t_2}}=-\frac{2}{3}\left(\matrix {-3v'_2\cr {u_{2}u'_{2}+\theta^{2}u'''_2} \cr}\right)
\end{equation}\\\\
The next step concerns the $\theta$-generalization of some properties, that we presented in an unpublished note [26], based essentially on the idea to use the dressing gauge group of Lorentz scalar momentum operators
$K[a]$ to discuss the Moyal DIZ-covariantization of ${sl_n}-{\widehat \Sigma}^{(0,n)}_{n}$ Lax operators and the primarity condition. \\\\
Recall that the standard method for constructing primary fields $w_k$, is based on the well known DIZ covariantization method of Lax differential operators [15]. This method has leads to build in a successful way all the $w_{k}$ conformal primary fields as functions of the old ones $u_k$ by covariantizing the corresponding Lax operators. Actually this method was extended and applied to the Moyal momentum case by the authors of [24] and has leads to build the $\theta$-deformation of the classical $w_k$-algebra. \\\\
Our alternative way consists in searching for the form of the dressing gauge symmetry $\{K[a]\}$ ensuring in one hand the transition 
\begin{equation}
\{u_i\}\stackrel{K[a]}{ \longrightarrow}\{V_i\}
\end{equation}
from different basis of conformal fields belonging to the ring ${\widehat \Sigma}_{n}^{(0,0)}, n\ge 0$ and on the primarity condition on the other hand. In fact, our request of the invariance of $sl_n$ momentum Lax operators under the action of the dressing gauge group makes strong constraints on the gauge parameters $a_i$ as it's explicitly shown in the $sl_3$ and $sl_4$ examples of sec.(5). Once these parameters are well derived, the associated gauge group is then explicitly determined. \\\\
In the same philosophy, the transition to a primary basis $w_k$ can be achieved via a dressing gauge group that we can explicitly determine. The knowledge of this gauge symmetry may provides a geometrical interpretation of the primarity condition of conformal fields as being a gauge choice on some orbit that we call the dressing gauge orbit. Elements of this orbit are well defined gauge dressing groups $\{K_{i}[a]\}$ such that each position on this orbit is characterized by a fixed dressing operator $K_{i}$ which fixes in turn the gauge group. \\\\
Throughout the computations that we performed, we selected some remarks that we present here: \\ \\
{\bf 1}. Some parameters of the dressing gauge group show a globally defined behavior (integral form). One can circumvent this property, originated from the operator character of the star product, by ignoring the non linear terms, which means also, doing a linear truncation on the dressing gauge group.\\\\
But this way to proceed makes the gauge symmetry partially defined. In order to avoid these kind of linear truncations, one has to consider from the beginning that the dressing group $\{K[a]\}$ ensuring the transition between deferents basis as a finite dimensional set, where only a finite number of the parameters $a_k$ which are well defined. This assumption is actually  compatible with the  derived results that we presented in the explicit examples of sec.(5).\\\\
{\bf 2}. As signaled in our last work [25], one can eventually consider fractional momentum algebra based on objects type $p^{\frac{a}{b}}$ and whose Leibnitz rule is shown to take the following form

\begin{equation}
{p^{\frac{a}{b}} \star f(x,p)} = {\sum _{s=0}^{\infty}}{\Pi _{j=0}^{s-1}}({\frac{a}{b}}-j) {\frac {\theta ^{s}}{s!}}f^{(s)}(x,p) p^{\frac{a}{b}-s},  
\end{equation}
This is a more general Moyal algebra as it can describes, formally, fractional spin objects and from which one can also recover all the known properties of the Das-Popowicz Mm algebra based on integer powers of the momenta. \\\\
It would be very interesting to look for the contribution and the meaning of these fractional powers of momenta in the framework of non-commutative KdV-hierarchies and also in the field theory building. This and other important points will be considered in our forthcoming works.
\\\\\\
{\bf Acknowledgements} \\ 
We would like to thank the Abdus Salam International Centre for Theoretical Physics (ICTP) for good hospitality and acknowledge the considerable help of the High Energy Section. We would like also to thank the associate-ship office and the Young Collaborator Program. This work was done at the high energy section within the framework of the associate-ship Scheme. We are also grateful to Ashok Das and Ziemowit Popowicz for their interest in this study. A. B. would like to thank the UFR-Faible Radiocativite, Physique mathematique et environnement, Departement de Physique, kenitra.

\newpage
{\bf References} 
\begin{enumerate} 
\item[[1]] For reviews see for instance;\\
L.D. Faddeev and L.A. Takhtajan, Hamiltonian Methods and the theory of solitons, Springer, 1987,\\
A. Das, Integrable Models, World scientific, 1989 and references therein.
\item[[2]] A.A. Belavin, A.M. Polyakov and A.B. Zamolodchikov, Nucl Phys. B241 (1984) 333-380;\\
V. S. Dotsenko and V.A. Fateev, Nucl Phys. B240 [FS12] (1984) 312-348;\\
P. Ginsparg, Applied Conformal field Theory, Les houches Lectures (1988);
\item[[3]] B.A. Kupershmidt, Phys. Lett. A102(1984)213;\\
Y.I. Manin and A.O. Radul, Comm. Math. Phys.98(1985)65;\\
\item[[4]] P. Mathieu, J. Math. Phys. 29(1988)2499;\\
W.Oevel and Z. Popowicz, Comm. math. Phys. 139(1991)441;
\item[[5]] J.C.Brunelli and A. Das, Phys. Lett.B337 (1994)303;\\
J.C.Brunelli and A. Das, Int. Jour. Mod. Phys. A10(1995)4563;
\item[[6]] E.H. Saidi and M.B. Sedra, Class. Quant. Grav.10(1993)1937-1946;\\
E.H. Saidi and M.B. Sedra, Int. Jour. Mod. Phys. A9(1994)891-913;
\item [[7]] E.H. Saidi and M.B. Sedra, J. Math. Phys. 35(1994)3190;\\
M.B. Sedra, J. Math. Phys. 37(1996)3483;
\item[[8]] A.B. Zamolodchikov, Teo. Math. Fiz.65(1985)374;\\
V.A.Fateev and S. Lukyanov, Int. Jour. Mod. Phys. A (1988)507.
\item[[9]] For a review see,\\
P. Bouwknegt and K. Schoutens, Phys. Rep. {\bf 223} (1993)183. and references theirein.
\item[[10]] A. Bilal and J.L. Gervais, Phys. lett. B 206 (1988) 412; Nucl. Phys. B 314 (1989)597;\\
I. Bakas, Nucl. Phys. B 302 (1988)189.
\item[[11]] J.L. Gervais, Phys. lett. B 160 (1985)277;
\item[[12]] P. Mathieu,  Phys. lett. B 208 (1988)101;
\item[[13]] K. Yamagishi, Phys. lett. B 259 (1991)436;\\
F. Yu and Y.S Wu, Phys. lett. B 236 (1991)220;\\
A. Das, W.J. Huang and S. Panda, Phys. lett. B 271 (1991)109;\\
A. Das, E. Sezgin and S. J. Sin, Phys. lett. B 277 (1992)435;
\item[[14]] I.M. Gel'fand and V. Sokolov, J. Sov. Math.30 (1985) 1975; Funkt. Anal. Priloz.10 (1976) 13; 13(1979)13.
\item[[15]] P. Di-Francessco, C. Itzykson and J.B. Zuber, Comm. Math. Phys. 140, 543 (1991)
\item[[16]] A. Connes, Noncommutative geometry, Academic Press (1994),\\
 A. Connes, M.R. Douglas, A. Schwarz, JHEP 02(1998) 003, [hep-th/9711162] and references therein.
\item[[17]] N. Seiberg and E. Witten, JHEP 09(1999) 032.
\item[[18]] H. Groenewold, Physica 12(1946)405,
\item[[19]] J.E. Moyal, Proc. Cambridge Phil. Soc. 45(1949)90,\\
M. Kontsevitch, [q-alg/9709040]\
\item[[20]] D.B. Fairlie, [hep-th/9806198],\\
D.B. Fairlie, Mod.Phys.Lett. A13 (1998) 263-274, [hep-th/9707190],\\
C. Zachos,[hep-th/0008010],\\
C. Zachos, J.Math.Phys. 41 (2000) 5129-5134, [hep-th/9912238],\\
C. Zachos, T. Curtright, Prog.Theor. Phys. Suppl. 135 (1999) 244-258, [hep-th/9903254].
\item [[21]] A. Das and Z. Popowicz, Phys.Lett. B510 (2001) 264-270, [hep-th/0103063].\
\item [[22]] A. Das and Z. Popowicz, J. Phys. A, Math. Gen.34(2001)6105-6117 and [hep-th/0104191].
\
\item [[23]] Ming-Hsien Tu, Phys.Lett. B508 (2001) 173-183, \ 
\item[[24]] Ming-Hsien Tu, Niann-Chern Lee, Yu-Tung Chen,  J.Phys. A35 (2002) 4375.\
\item [[25]] A. Boulahoual and M.B. Sedra; hep-th/0207242, submitted to Mod. Phys. Lett. A\
\item[[26]] M. Rachidi, E. H. Saidi and M. B. Sedra, ICTP preprint, IC/95/176;\
\item[[27]] J. Boussinesq, Comptes Rendus, 1871, V.72, 755-759,\\
M. Ablowitz, H. Segur, ``Solitons and the Inverse Scattering Transform'' SIAM Philadelphia 1981.\
\end{enumerate} 
\end{document}